\documentclass[aps,prd,twocolumn,eqsecnum,amssymb,amsmath,showpacs,a4paper, superscriptaddress]{revtex4-2}

\usepackage{bm}
\usepackage{hyperref}
\usepackage{framed}
\usepackage{subfigure}
\usepackage{hyperref}   % [breaklinks,colorlinks]
\usepackage{graphicx}
\usepackage{amsmath}
\usepackage{color}
\usepackage{yfonts}

\usepackage{hyperref}
\hypersetup{
    colorlinks=true,
    linkcolor=blue,
    filecolor=magenta,
    urlcolor=cyan,
}
\newcommand{\grad}{\bm{\nabla}}
\newcommand{\n}{\bar{n}}
\newcommand{\Lin}{\mathcal{L}}

\newcommand{\vor}{\mathrm{vor}}
\newcommand{\ph}{\mathrm{ph}}
\newcommand{\coef}{\rho}

\begin{document}

\title{Rabi oscillations with close-range quantum vortex states}

\author{Sam Patrick}
\email{samuel\_christian.patrick@kcl.ac.uk}
\affiliation{Department of Physics, King’s College London, University of London, Strand, London, WC2R 2LS, UK}

\date{\today}

\begin{abstract}
Quantum vortices separated through distances much larger than their core size interact via their long-range velocity field.
At smaller separations, however, the influence of the core's compressibility strongly influences the vortex dynamics.
Using the example of a compact ring of five vortices, it is shown that close-range effects lead to a new (slower) state of orbital motion which is not predicted by the usual long-range theory. 
This secondary state can be created starting from the usual orbital state by modulating the interaction strength to induce Rabi oscillations between the states.
\end{abstract}

\maketitle

\emph{Introduction.}---Vortices are key components of a wide variety of systems, examples of which include the ocean and atmosphere, optical devices, superconductors and neutron stars \cite{mcwilliams1985submesoscale,kuo1966dynamics,grier2003revolution,blatter1994vortices,wlazlowski2016vortex}.
% cosmology hindmarsh1995cosmic, non-abelian gauge theories tong2009quantum
In quantum fluids (such as atomic Bose-Einstein condensates (BECs) \cite{fetter2001vortices}, superfluid $^4$He \cite{barenghi2001quantized}, polariton condensates \cite{carusotto2013quantum}) vortices are topological defects of the system's order parameter $\Psi$ with quantised circulation $\oint d\bm{l}\cdot\mathbf{v} = 2\pi\hbar\ell/M$, where $\mathbf{v}=\grad\mathrm{arg}(\Psi)$ is the fluid velocity, $M$ is the mass of the condensate particles, and $\ell\in\mathbb{Z}$ is the number of times the phase of $\Psi$ winds around the defect.
The existence of quantum vortices (QVs) in these systems gives rise to important phenomena such as mutual friction \cite{barenghi2001quantized}, topological phase transitions \cite{ryzhov2017berezinskii} and quantum turbulence in 2D \cite{neely2013characteristics} and 3D \cite{kozik2009theory}. 

At low energy, the separation between QVs is much larger than the size of the vortex core.
Hence, the core structure is neglected and QVs are treated as infinitesimal filaments (in 3D) or points (in 2D) interacting only through their long-range velocity field.
In 2D, the corresponding model is known as the point-vortex (PV) model which, without dissipation, takes the form,
\begin{equation} \label{PVmodel}
	\dot{\mathbf{x}}_i = \sum_{j\neq i}\mathbf{v}_j(\mathbf{x}_i),
\end{equation}
where $\mathbf{v}_j$ is the velocity field of the $j^\mathrm{th}$ vortex whose centre is at the position $\mathbf{x}_j(t)$, overdot denotes time differentiation and a physical boundary can be accounted for by including image vortices.
Eq.~\eqref{PVmodel} states that each vortex simply moves with a velocity given by summing up the velocity fields of all the others.
Although the PV model captures the salient features of the vortex dynamics, it can fail quantitatively in simple scenarios where the fluid compressibility (density variations) cannot be neglected \cite{groszek2018motion}.
Furthermore, the model does not account for important phenomena arising for vortices at close-range, i.e. when the inter-vortex distance becomes small.
Indeed, Eq.~\eqref{PVmodel} has to be supplemented with ad hoc rules to describe elementary processes such as vortex mergers, annihilations and reconnections \cite{carnevale1991evolution,barenghi2001quantized}.
 
% \emph{Introduction.}--Vortices are often treated as point-like (2D) and lines of vanishing cross-sections (3D). This picture is useful for studying many phenomena, e.g. Kelvin waves, quantum turbulence and associated energy cascades, topological (BKT) phase transitions, neutron star glitches?, cosmic strings. It is most useful when the vortices can be considered dilute, i.e. Intervortex separation is larger than the size of the core. Say a bit about the approximation, interaction through the long-range velocity field, phase is the only important thing and density variations neglected, low-energy (hydrodynamic) theory. The compressibility of the core becomes important when the intervortex separation becomes comparable to the healing length. In dense vortex clusters, point-vortex model fails to make accurate quantitative predictions \cite{groszek2018motion}.  Important phenomena such as annihilation of vortex/anti-vortex pairs (2D), vortex reconnections and vanishing of vortex rings (3D), and splitting of multiply quantised vortices.

QV splitting instabilities are another crucial effect in this category, since they cause QVs with $|\ell|>1$ to decay into clusters of $|\ell|=1$ vortices.
There is a rich literature on QV splitting patterns, especially in the context of atomic BECs.
The simplest case involves the splitting of $\ell=2$, which was observed in \cite{shin2004dynamical}.
% Splitting patterns of $\ell=2,3$ QVs can be controlled by adjusting the interaction strength \cite{pu1999coherent} and splitting-merging revivals can occur due to finite size effects \cite{nilsen2008splitting,patrick2022origin}. Splitting of an $\ell=2$ QV due the dynamical instability was observed experimentally in \cite{shin2004dynamical}.
At higher $\ell$, the dynamics becomes much richer due to the competition of multiple unstable modes.
The $\ell=4$ QV (which is predicted to be unstable for $m=2,3,4$ \cite{kawaguchi2004splitting}) can be created in $^{87}$Rb by reversing magnetic field of the trap \cite{kumakura2006topological}.
Experiments on $\ell=4$ have measured the linear ($m=2$) and triangular ($m=3$) splitting patterns \cite{okano2007splitting,isoshima2007spontaneous,kuwamoto2010dynamics,shibayama2016density}, but not others due narrow instability windows and small growth rates in finite size set-ups.
The splitting dynamics is strongly influenced by the BEC interaction strength \cite{pu1999coherent} and trap geometry \cite{nilsen2008splitting,zhu2021splitting,patrick2022origin}.
In holographic superfluids (studied using the AdS/CFT correspondence) temperature is an additional control parameter \cite{lan2023splitting} and it has been argued that a temperature increase can favour the $m=4,5,6$ instability of the $\ell=4$ QV \cite{lan2023heating}.
It is also possible to prevent splitting using a mexican hat shaped trap \cite{adhikari2019stable}, inducing a convergent flow through particle depletion \cite{srinivasan2006vortices,alperin2021multiply,ruffenach2023superfluid,delhom2024entanglement,patrick2024quantum,svanvcara2024rotating}, modulating the interaction strength \cite{wang2011quantized} or adding a secondary particle species inside the vortex core \cite{patrick2023stability}.
Further novelties can be observed in two-component BECs when the species are miscible, including $\ell=1$ instabilities and splitting-merging revivals \cite{berti2023superradiant}, $\ell>1$ ground states \cite{kuopanportti2015ground} and temperature dependent dynamical transitions in composite vortices \cite{an2025splitting}.

Overall, there is a huge interest in understanding the splitting of large QVs and the ensuing dynamics.
In most cases discussed above, the vortex motion following the splitting is captured by Eq.~\eqref{PVmodel} at late times, at least in large systems.
The purpose of letter is to illustrate that a new class of vortex states (with no counterpart in the long-range PV theory) can arise for QVs at close-range, and that these states can be accessed starting from the usual states by modulating the interaction strength to induce Rabi oscillations.

\emph{Vortex states.}---We consider the (dissipative) Gross-Pitaevskii equation (GPE), 
\begin{equation} \label{GPE}
i\hbar\partial_t\Psi = (1-i\gamma)\left(-\frac{\hbar^2}{2M}\nabla^2+ V(\mathbf{x}) + g|\Psi|^2 -\mu\right)\Psi,
\end{equation}
describing the condensed fraction of a cold atomic gas, e.g $^{23}$Na or $^{87}$Rb \cite{shin2004dynamical,okano2007splitting}, near $T=0\,\mathrm{K}$.
Variants of the GPE describe polariton condensates and quantum fluids of light \cite{carusotto2013quantum}.
We assume a quasi-2D condensate in the $\mathbf{x}=(x,y)$ plane, with tight confinement in the vertical direction.
$V(\mathbf{x})$ is the trapping potential, $g$ is the interaction strength, $\mu$ is the chemical potential and $\gamma$ is a phenomenological dissipation parameter \cite{cockburn2009stochastic}.
The system is described by the condensate wavefunction $\Psi=\sqrt{\n}e^{i\Phi}$, where $\n$ is the number density of atoms in the condensate which move with velocity $\mathbf{v} = \frac{\hbar}{M}\grad\Phi$.
The ratio of the nonlinear term to the kinetic term defines the healing length $\xi=\hbar/\sqrt{M\mu}$.
The speed of sound (density waves) in the uniform state $\n=\mu/g$ is given by $c = \sqrt{\mu/M}$.
In what follows we set $\hbar,M,\mu=1$, which amounts to measuring distance and time in units of $\xi$ and $\xi/c$ respectively.

Eq.~\eqref{GPE} has stationary vortex solutions given by $\Phi=\ell\theta$, where $\ell$ is the winding number mentioned in the introduction and $\mathbf{x}=(r,\theta)$ are polar coordinates. We fix $\ell>0$ without loss of generality and consider a vortex at the centre of a circular trapping potential with a hard wall at $r=r_B$. The density $\n(r)$ (determined numerically by evolving \eqref{GPE} in imaginary time) vanishes exponentially outside $r_B$ and scales as $\n{\sim}r^{2\ell}$ inside the vortex core, which has characteristic width $\ell\xi$.
When the vortex is excited, the full order parameter becomes $\Psi=e^{i\ell\theta}(\sqrt{\n}+\psi)$. 
Small excitations can be expanded in modes $\psi = \sum_m (u_m e^{im\theta} + v_m^*e^{-im\theta})$ with $m\in(-\infty,\infty)$, where $u_m$ and $v^*_m$ are the positive and negative frequency components of $\psi$ respectively. These obey the linear equation,
\begin{equation} \label{Lin1}
\begin{split}
& i\partial_t|\psi_m\rangle = \Lin|\psi_m\rangle, \quad \Lin = \begin{pmatrix}
\mathcal{D}_+ & g\n \\ -g\n & -\mathcal{D}_-
\end{pmatrix}, \\
& \mathcal{D}_\pm = -\frac{1}{2}\nabla^2_r + \frac{(\ell\pm m)^2}{2r^2}+2g\n+V-1,
\end{split}
\end{equation}
where $|\psi_m\rangle = (u_m, v_m)^\mathrm{T}$.
In the stationary state, $|\psi_{m}\rangle = \sum_n a_n e^{-i\omega_n t}|\psi_{mn}\rangle$ where $a_n$ are the mode amplitudes and $n\in[0,\infty)$ indexes the different eigenmodes.
These are normalised in the inner product,
\begin{equation}
\langle \psi_\lambda | \psi_{\lambda'} \rangle_{\sigma_3} \equiv \int d^2\mathbf{x}(u_\lambda^* \ v_\lambda^*)\sigma_3\begin{pmatrix}
u_{\lambda'} \\ v_{\lambda'}
\end{pmatrix} = \mathcal{N}_\lambda\delta_{\lambda\lambda'},
\end{equation}
where $\sigma_3$ is the third Pauli matrix, $\mathcal{N}_\lambda = \pm 1$ is the sign of the norm, and we write $\lambda=(m,n)$ for shorthand.
The eigenfrequencies $\omega_\lambda$ are determined by the eigenvalue equation $\omega_\lambda|\psi_\lambda\rangle = \Lin|\psi_\lambda\rangle$ and fall into two categories.
Modes which are localised in the vortex core (with $\omega^\vor$) encode the vortex dynamics, whereas those which extend outside the vortex core are sound waves (phonons with $\omega^\ph$).
If $\omega^\vor$ and $\omega^\ph$ are sufficiently close, they hybridise to produce a dynamical instability describing the transfer of energy between the central vortex and a surrounding phonon \cite{patrick2022quantum,patrick2022origin}.
By selecting a small value of $r_B$, this instability can be avoided since $\omega^\ph\sim r_B^{-1}$ at low frequency whilst $\omega^\mathrm{vor}$ is roughly $r_B$ independent.

\begin{figure*} %[!h] 
\centering
\includegraphics[width=\linewidth]{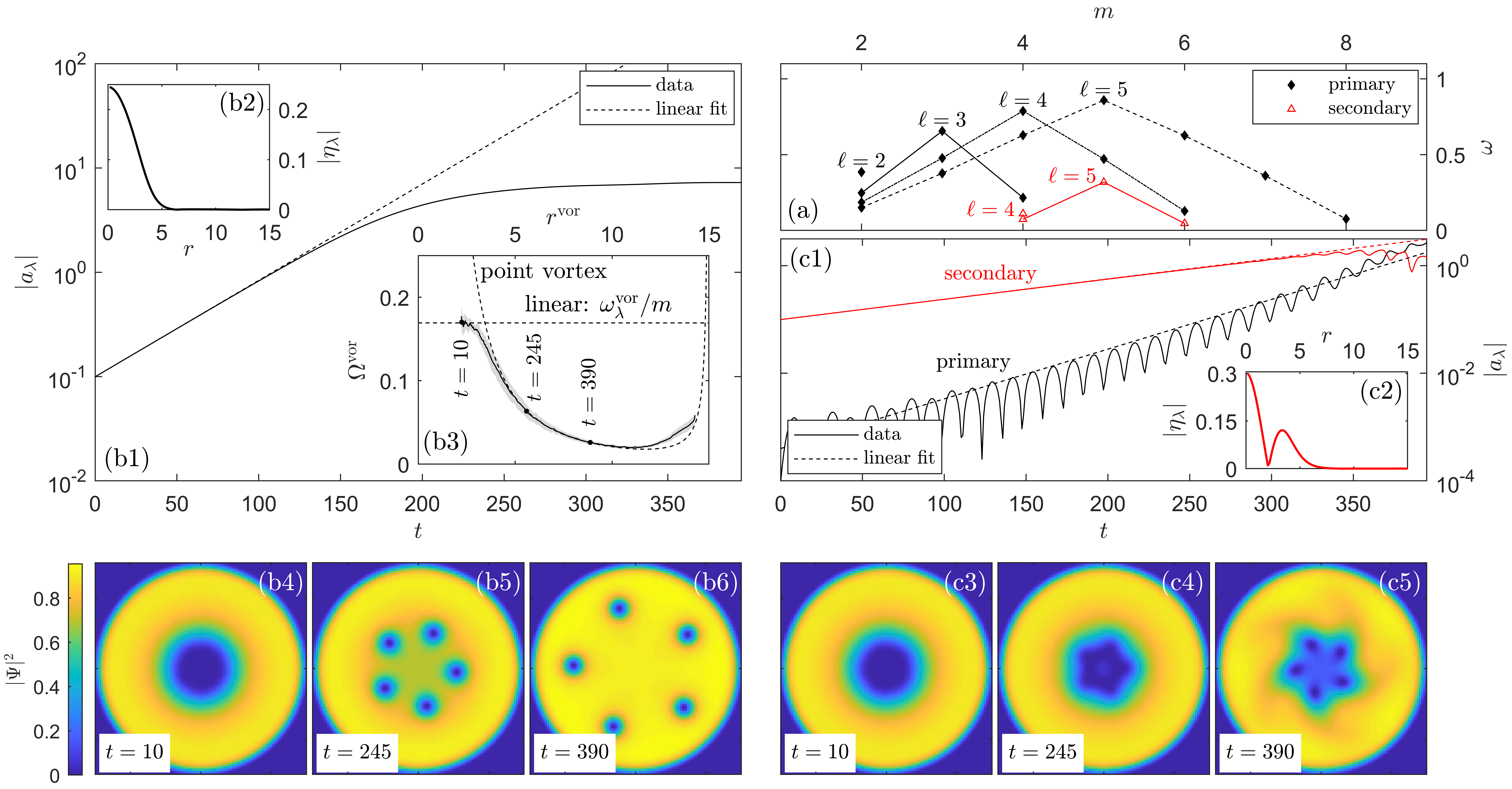}
\caption{\textbf{(a)}: Frequency of the vortex excitations computed using the WKB method of \cite{patrick2022quantum}. \textbf{(b)}: Evolution of the $m=5$ primary mode around an $\ell=5$ vortex. The trapping potential is $V=V_0\big[1+(V_0-1)e^{a(r_B-r)}\big]^{-1}$ with $V_0=a=10$ and $r_B=15$. (b1) The primary mode amplitude (solid line) agrees with the linear growth rate (dashed line) at early times. (b2) Representative mode function $\eta_\lambda=u_\lambda+v_\lambda$ for the primary mode. (b3) The orbital frequency $\Omega^\vor$ of the 5 vortices (black line) compared with the linear theory and PV predictions (uncertainty in vortex tracking shown in grey). (b4-b6) $|\Psi|^2$ at the three different times shown as black points on panel (b3). \textbf{(c)}: The same for a seed secondary mode. (c1) Primary and secondary mode growth rates (black and red solid lines) agree with the linear growth rates (dashed lines). (c2) Mode function of the secondary mode. (c3-c4) Growth of the secondary mode does not split the vortices but leads to an arrangement with overlapping vortex cores. (c5) The vortices split when the primary mode becomes dominant. 
} \label{fig:1}
\end{figure*}

In the following, we will be primarily interested in the dynamics of the vortex modes.
Those with $m>0$ have $\mathcal{N}_\lambda = -1$ and describe negative energy excitations, that is, states which lower the total energy of the vortex. 
The different $\mathcal{N}_\lambda = -1$ states correspond to distinct configurations of vortices as the cluster loses energy.
They are approximately determined (when the density varies smoothly) by a resonance condition $\cos S(\omega^\vor_\lambda) = 0$, where $S$ is the phase integral of the mode over the vortex core (see \cite{patrick2022quantum,patrick2022origin} for details). $S$ increases with the size of the vortex core, hence there will be an increasing number of states for larger $\ell$. In Fig.~\ref{fig:1}(a), we plot the zeros of $\cos S(\omega^\vor_\lambda)$ for $\ell=2,3,4,5$.
In general, there are negative energy vortex modes in the band $2\leq m\leq 2\ell-2$ \cite{giacomelli2020ergoregion}.
For $\ell=2$, the only such mode occurs for $m=2$ and encodes the splitting into two vortices with $\ell=1$, whereas for $\ell=3$, there are states for $m=2,3,4$, corresponding to a linear/triangular/square arrangement of $\ell=1$ vortices, where the latter also includes an $\ell=-1$ vortex at $r=0$.
A new phenomenon occurs for $\ell=4$, namely, the occurrence of a secondary (lower frequency) mode with $m=4$, in addition to the primary mode present at lower $\ell$.
These secondary modes become more numerous for larger $\ell$, e.g. $\ell=5$ possesses secondary modes for $m=4,5,6$.
From here on, we focus on the case $\ell=m=5$ due to numerical artefacts when working with a fourfold symmetric Cartesian grid.

The secondary modes correspond to a distinct state of orbital motion with no counterpart in the PV dynamics.
To verify this, we simulate \eqref{GPE} with an initial condition containing an $\ell=5$ vortex seeded with a mode amplitude $a_{\lambda_i}=0.1$ in either the primary ($\lambda_p$) or secondary ($\lambda_s$) vortex mode.
The numerical method is based on the split-step Fourier method described in \cite{patrick2022origin}.
For the interaction parameter, we take $g=1$ which amounts to a rescaling of $\Psi\to g^{-1/2}\Psi$.
A small amount of damping ($\gamma=2.5\times 10^{-2}$) causes the amplitude of the seed mode to grow.
See \footnote{Simulations were performed on a square grid $x\in[-L,L]$, with $L=r_B+5$ and $N_x=128$ grid points (similarly for $y$). The time step was $\Delta t=L^2/5N_x^2$. The initial vortex profile $\sqrt{\n}e^{i\ell\theta}$ was found by evolving Eq.~\eqref{GPE} in imaginary time by factoring out the phase and simulating only the radial direction. Eigenmodes are obtained by diagonalising $\omega_\lambda|\psi_\lambda\rangle = \Lin|\psi_\lambda\rangle$, approximating the radial derivatives with 5-point centred finite difference stencils everywhere except $r=0$ ($r=L$) where we apply a 5-point forward (backward) stencil and a regularity (Dirichlet) boundary condition.} for numerical details.

When the initial condition contains the primary mode, the exponential growth makes the central vortex split into a vortex ring, whose orbital frequency $\Omega^\mathrm{vor}$ coincides with the PV model at large vortex separation (see Fig.~\ref{fig:1}(b)).
Since the PV orbital frequency smoothly merges with $\omega^\vor_{\lambda_p}/m$, we deduce that the primary mode is the extension of the dynamics described by Eq.~\eqref{PVmodel} when the vortex separation goes to zero (this is also seen for $\ell=m=2$ in \cite{patrick2022origin}).
Conversely, a seed secondary mode does not cause splitting and the vortex cores remain overlapping (see Fig.~\ref{fig:1}(c)).
This difference is due to the different mode functions shown in panels (b1) and (c1).
The primary mode has a single extremum at $r=0$, which causes condensate particles to fill in the centre and the vortices to move out into a ring.
Conversely, the secondary mode has two extrema, and the zero between them means the $\Psi$ retains its background value there, preventing the vortices from separating. Eventually, the amplitude of the primary mode (which grows faster) dominates and the vortex splits.
Any attempt to push the secondary mode to larger vortex separations (i.e. increasing the seed amplitude $a_{\lambda_s}$) excites the primary mode and leads to splitting.
The secondary mode appears, therefore, to be a unique feature of close-range vortex dynamics without a counterpart in the long-range (PV) theory.

\begin{figure*} %[!h] 
\centering
\includegraphics[width=.95\linewidth]{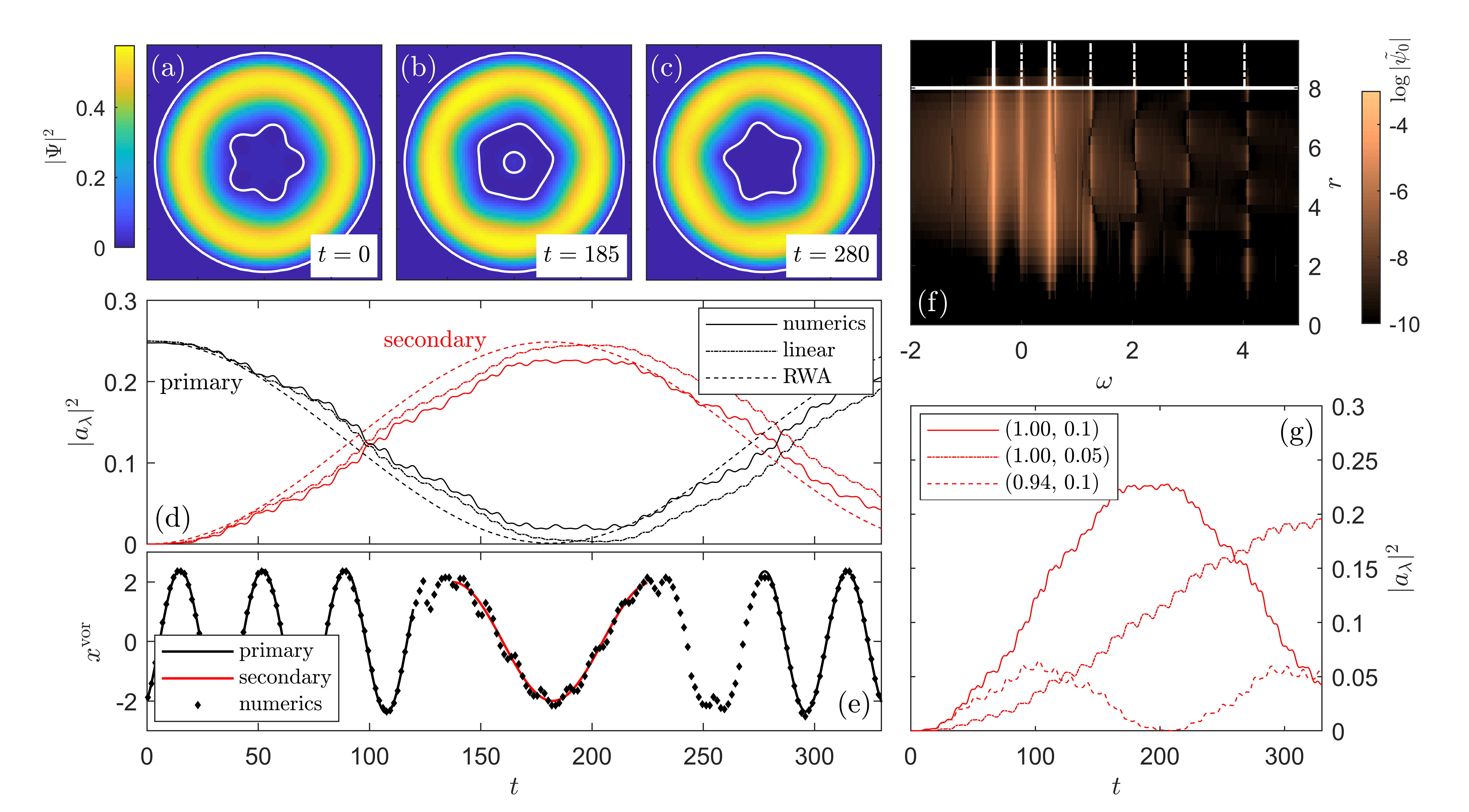}
\caption{\textbf{(a-c)}: $|\Psi|^2$ at three representative times showing the density profile of the primary (a), the secondary (b) and a mixture of the two (c). \textbf{(d)}: The mode amplitudes during the simulation as a function of time, compared against the linear model in Eq.~\eqref{AmpEq2} and the RWA prediction of Eq.~\eqref{RWAsol}. \textbf{(e)}: The $x$ coordinate of one of the vortices, fit with the frequency $\omega_{\lambda_i}/m$ for the primary ($\lambda_p$) and secondary ($\lambda_s$) mode. \textbf{(f)}: Fourier transform of the $m=0$ excitation induced by varying the interaction parameter compared against the modulation frequency $\Omega$ (solid vertical white lines) and the $m=0$ eigenmodes (dash-dotted vertical white lines). \textbf{(g)}: Secondary mode amplitudes for different values of $(\zeta,\delta g_0)$ shown in the legend.
} \label{fig:2}
\end{figure*}

\emph{Exciting the secondary mode.}---Since the primary mode is more unstable, it is natural to wonder whether there is any realistic scenario where the secondary mode can be observed. We show now that the secondary mode can be accessed starting from the primary by modulating one of the system parameters at a frequency $\Omega=\zeta(\omega^\vor_{\lambda_p}-\omega^\vor_{\lambda_s})$, where $\zeta$ is close to $1$, to induce Rabi oscillations between the states. We choose to vary the interaction parameter,
\begin{equation} \label{goft}
g(t) = 1+\delta g, \qquad \delta g = \delta g_0 \sin\Omega t,
\end{equation}
which could be achieved with a Feshbach resonance \cite{timmermans1999feshbach}. Another option would be to work with a harmonic trapping potential and modulate the trap frequency.
The following treatment shares similarities with \cite{caradoc1999coherent}, where a periodic stirring potential was used to drive nonlinear transitions between the ground state and single vortex state.

The first effect of this modulation is to alter background (symmetric) vortex state.
Provided $|\delta g_0|\ll 1$, this shift can be treated as a linear perturbation which obeys,
\begin{equation}
i\partial_t|\psi_0\rangle = \Lin|\psi_0\rangle + \delta g(t) \n^{3/2}\begin{pmatrix}
1 & -1
\end{pmatrix}^\mathrm{T},
\end{equation}
with $|\psi_0\rangle = (\psi_0\ \psi_0^*)^\mathrm{T}$ and $\psi_0=\int^{2\pi}_0\frac{d\theta}{2\pi}(\Psi e^{-i\ell\theta}-\sqrt{\n})$.
This $\psi_0$ will modulate the parameters in the $m\neq 0$ equation, which to linear order in $\delta g_0$ is,
\begin{equation} \label{Lin2}
\begin{split}
& i\partial_t|\psi_m\rangle = (\Lin+\delta\Lin)|\psi_m\rangle, \quad \delta\Lin = \begin{pmatrix}
A & B \\ -B^* & -A^*
\end{pmatrix}, \\
& A = 2\n\delta g + 4\sqrt{\n}\mathrm{Re}[\psi_0], \quad B = \n\delta g + 2\sqrt{\n}\psi_0.
\end{split}
\end{equation}
All the $\omega_\lambda$ are assumed real, which occurs provided we don't have a dynamical instability (i.e. for small trap sizes as explained above).
Applying basic techniques, the state is projected onto the basis of the linear equations $|\psi_m\rangle = \sum_n a_\lambda(t)e^{-i\omega_\lambda t}|\psi_\lambda \rangle$ with amplitudes $a_\lambda$ which vary slowly relative to $\omega_\lambda^{-1}$ (recall that $\lambda=(m,n)$).
Since the background is symmetric in $\theta$ and Eq.~\eqref{Lin2} is linear, the different $m$ do not couple.
We therefore work with a single mode ($m=5$ in our case) and drop the index $m$.
The equation for each mode amplitude becomes,
\begin{equation} \label{AmpEq1}
i\mathcal{N}_n\dot{a}_n = \sum_{n'} \coef_{nn'}(t) e^{i(\omega_n-\omega_{n'})t},
\end{equation}
where $\coef_{nn'}(t) = \langle \psi_n|\delta\Lin|\psi_{n'}\rangle_{\sigma_3}$.
We now make the simplifying assumption that the system behaves as a two level system, which is satisfied in the usual way provided the background modulation $\Omega$ is close to the level difference.
We take the two levels to be the primary and secondary vortex modes, which for small traps ($r_B\lesssim 10.5$) correspond to $n=1$ and $n=0$ respectively.
If modulations in $g$ are switched on abruptly, $\delta\Lin$ will contain multiple frequencies. 
There will be a dominant component at $\omega=\pm\Omega$ and subdominant components at the $m=0$ eigenfrequecies, provided $\Omega$ is not on resonance with these. 
Focussing on only the dominant contributions, we can approximate $\coef_{nn'} \simeq \tilde{\coef}^+_{nn'}e^{-i\Omega t} +\tilde{\coef}^-_{nn'}e^{i\Omega t}$, where $\tilde{\coef}^\pm_{nn'}$ are the Fourier components of $\coef_{nn'}$ at $\omega=\pm\Omega$.
Writing $a_{n}(t) = b_n(t)\exp(i\int_0^t \coef_{nn}(t')dt')$, the $b_n$ obey,
\begin{equation} \label{AmpEq2}
\begin{split}
i\dot{b}_0 = & \ -\tilde{\coef}^+_{01}e^{-i\int_0^t\Delta^+ dt'}b_1 -\tilde{\coef}^-_{01}e^{-i\int_0^t\Delta^-dt'}b_1, \\
i\dot{b}_1 = & \ -\tilde{\coef}^+_{10}e^{i\int_0^t\Delta^- dt'}b_0 -\tilde{\coef}^-_{10}e^{i\int_0^t\Delta^+dt'}b_0,
\end{split}
\end{equation}
with $\Delta^\pm = \omega_1-\omega_0-\coef_{11}+\coef_{00}\pm\Omega$.
The terms $\coef_{00}$ and $\coef_{11}$ are oscillatory and small compared to the remaining terms which lead to secular evolution of the phase.
Neglecting them, the $\Delta^\pm$ become constants.
Furthermore, in the rotating wave approximation (RWA), the terms containing $\Delta^+$ oscillate much more rapidly than the ones with $\Delta^-$ and tend to average out.
Hence, writing $\tilde{\coef}\equiv\tilde{\coef}^-_{01}=\tilde{\coef}^{+*}_{10}$ and $\Delta\equiv\Delta^-$, the equations become,
\begin{equation} \label{RWAeq}
i\dot{b}_0 = -\tilde{\coef} e^{-i\Delta t}b_1, \qquad i\dot{b}_1 = -\tilde{\coef}^* e^{i\Delta t}b_0.
\end{equation}
The solution for an initial amplitude $b_1(0)=\alpha$ in the primary mode and $b_0(0)=0$ in the secondary is,
\begin{equation} \label{RWAsol}
\begin{split}
b_1(t) = & \ e^{\frac{i\Delta t}{2}}\alpha\left(\cos\kappa t -\frac{i\Delta}{2\kappa}\sin\kappa t\right), \\
b_0(t) = & \ e^{-\frac{i\Delta t}{2}}\frac{i\tilde{\coef}\alpha}{\kappa}\sin\kappa t, \quad \kappa = \frac{\sqrt{\Delta^2+4|\tilde{\coef}|^2}}{2}.
\end{split}
\end{equation}
These describe out-of-phase oscillations with period $2\pi/\kappa$ of the mode occupation numbers.
Using $\Omega = \zeta(\omega_1-\omega_0)$, we have $\Delta = (\zeta-1)(\omega_1-\omega_0)$.
Hence, when the parameter $\zeta$ is close to 1, the primary mode amplitude $|b_1|$ can be brought close to zero at $t=\pi/2\kappa$, where the secondary mode amplitude $|b_0|$ reaches a maximum of around $|\alpha|$.

\emph{Numerical simulations.}---To verify the predictions of \eqref{RWAsol}, we perform a numerical simulation of \eqref{GPE} for the parameters $\gamma=0$ and $r_B=8$. In the initial condition, we seed the stationary $\ell=5$ vortex with the $m=5$ primary mode of amplitude $\alpha=0.5$. For the interaction parameter in \eqref{goft}, we use the values $\delta g_0=0.1$ and $\zeta = 1.00$.
Snapshots of the evolution are shown in Fig.~\ref{fig:2}(a-c), with contours of $|\Psi|^2=0.02$ shown in white. Panel (a), where vortex is in the primary state, has a distinct pattern from the panel (b) where the secondary mode is excited.

The mode amplitudes $a_\lambda$ during the simulation are computed by projecting $\Psi$ onto the eigenbasis $|\psi_\lambda\rangle$ \footnote{The seeding of a linear mode in the nonlinear equation \eqref{GPE} introduces a correction to the vortex density due to backreaction effects. This induces a small shift in the chemical potential $\mu$ which is calculated by measuring the phase drift in $\Psi$ over the duration of the simulation. Using the corrected chemical potential, the normal modes of the vortex are recomputed.	 This procedure slightly shifts the $\omega_\lambda$, hence why $\zeta$ is quoted to 2 decimal places.}.
We display $|a_\lambda|^2$ for the primary and secondary vortex modes in Fig.~\ref{fig:2}(d), observing that the minimum (maximum) of the primary (secondary) mode amplitude coincides with the distinct density pattern in panel (b).
The amplitudes are in good agreement with the predictions of the linear model in Eq.~\eqref{AmpEq1}, with differences expected due to the model not accounting for nonlinear mode interactions.
The two states also correspond to different vortex orbital motion, which we verify by tracking the vortices during the simulation.
In Fig.~\ref{fig:2}(e), the $x$-coordinate of one of the vortices is plotted (black diamonds) and compared a sinusoidal oscillation $\Omega^\vor$ corresponding to the primary mode (black line) and secondary mode (red line).
At the beginning and end of the simulation, the vortices are in the fast orbital state, corresponding to the close-range modification of the PV dynamics in Eq.~\eqref{PVmodel}.
At intermediate times, the vortices orbit more slowly, which is only possible due to the overlapping vortex cores.

In Fig.~\ref{fig:2}(e), we show the frequency content of the $m=0$ mode, which is extracted by computing $\tilde{\psi}_0(\omega,r)=\int dt \,e^{i\omega t}\int^{2\pi}_0\frac{d\theta}{2\pi}(\Psi e^{-i\ell\theta}-\sqrt{\n})$.
This is compared with the modulation frequency bands at $\pm\Omega$ (solid vertical white lines) and the $m=0$ eigenfrequencies (dashed-dotted certical white lines).
Noting that the dominant and leading sub-dominant contributions come from $\pm\Omega$ and the lowest phonon frequencies at $\pm\omega_0$ (ignoring the $\omega=0$ mode), we fit the coupling coefficients $\coef_{nn'}$ to a sum of four sinusoids at $\pm\Omega,\pm\omega_0$ to extract the Fourier amplitudes $\tilde{\coef}^\pm_{nn'}$ used in Eq.~\eqref{AmpEq2}.
The dominance of the $\pm\Omega$ justifies the approximations leading to the RWA results in Eq.~\eqref{RWAsol}.
These are also plotted in Fig.~\ref{fig:2}(d) and show good qualitative agreement with the numerical results. 
Despite neglecting both the coupling to $m=5$ phonons and the $m=0$ eigenfrequencies in $\coef_{nn'}$, the RWA result gives a good estimate of $\kappa$.

Finally in Fig.~\ref{fig:2}(g), we illustrate the effect of changing the modulation frequency $\Omega$ and amplitude $\delta g_0$, showing only the amplitude of the secondary mode. As expected from the RWA result, the secondary mode undershoots its maximum amplitude when $\Omega$ is detuned from $\omega_1-\omega_0$, whilst a smaller value of $\delta g_0$ leads to a decrease in $\kappa$.
Crucially, optimal conversion of the primary into the secondary mode is achieved when $\Omega$ matches the level difference $\omega_1-\omega_0$.

\emph{Discussion.}---We have demonstrated the existence of new (secondary) states of vortex motion not captured by the point-vortex model in Eq.~\eqref{PVmodel}. It was shown that these states can be excited by modulating the interaction strength to induce Rabi oscillations starting from the usual (primary) states of vortex motion, which are extensions of the dynamics in Eq.~\eqref{PVmodel} to small inter-vortex distances.
The main feature which led to these new states is the widened vortex core at large winding numbers $|\ell|$.
These states may be relevant for the dynamics of larger vortices in a variety of scenarios beyond the simple Gross-Pitaevskii model we have studied here, e.g. thin superfluid $^4$He films \cite{sachkou2019coherent}, exciton-polariton condensates \cite{alperin2021multiply}, intertype superconductors \cite{vagov2016superconductivity} and superconductor-ferromagnet structures \cite{aladyshkin2009nucleation}.

In atomic BECs, the splitting patterns which have received most attention are the ones corresponding to the primary states.
By contrast, the secondary states (which have lower frequencies) seem to be a consequence of overlapping vortex cores and do not extend to the regime of large vortex separations.
For giant vortices (with even larger $\ell$) an increasing number of these modes per $m$ can fit inside the vortex core.
A better understanding of them could therefore lead to improved modelling of compact vortex clusters in a wide variety of systems \cite{stockdale2020universal,alperin2021multiply,geelmuyden2022sound,delhom2024entanglement,patrick2024quantum,svanvcara2024rotating}. 

Whilst we have focussed on $\ell=5$ (for numerical purposes), the secondary mode first appears for $\ell=4$ with $m=4$ and  $\omega\sim 0.1 c/\xi$.
$\ell=4$ vortices have been the subject of various experimental studies \cite{okano2007splitting,isoshima2007spontaneous,kuwamoto2010dynamics,shibayama2016density}, firstly, because they are large enough that the competition of multiple unstable modes leads to rich dynamics and, secondly, since they can be induced by reversing the trap's magnetic field \cite{kumakura2006topological}.
Future experiments could therefore study these secondary states using the Rabi oscillation technique suggested here, either by modulating the interaction strength with a Feshbach resonance or the trap size.
It is feasible that a similar Rabi oscillation technique could also induce transitions between the threefold and fourfold symmetric splitting patterns of the $\ell=4$ vortex, where the latter is difficult to access due to its slow decay \cite{shibayama2016density,lan2023heating}.
Secondary states may also be relevant for dynamics of vortex filaments in 3D, for example, in the turbulent cascade induced by the decay of $\ell=4$ vortices \cite{cidrim2017vinen}.

Close-range effects may also be relevant for the statistical behaviour of large numbers of vortices -- so-called vortex gases.
In kinetic theories of vortices, the system can become frozen in a metastable state before reaching statistical equilibrium if only two-body collisions are considered, and collisions involving $\geq3$ vortices are required to unfreeze the system on longer timescales \cite{chavanis2001kinetic,chavanis2012kinetic2}.
Furthermore, three-body collisions are known to play an important role in the asymptotic scaling of the vortex density when it becomes small at late times \cite{sire2000numerical,sire2011effective}, highlighting the importance of collisions between multiple vortices.
These studies, typically based on the PV model, need to implement ad hoc rules which take effect at close-range \cite{carnevale1991evolution}.
In quantum fluids, vortex clustering occurs when suitable forcing, e.g. mechanical stirring, is used to overcome the effects of dipole decay \cite{reeves2013inverse,gauthier2019giant}, in which case the inverse cascade can become the dominant mechanism in energy transport \cite{bradley2012energy}, even leading to the formation of point-like giant vortex structures in the high energy limit \cite{yu2016theory}.
Compact clusters are also relevant when there is a large vortex in the initial condition \cite{stockdale2020universal}, and for quantum fluids with localised heating \cite{sachkou2019coherent} or particle losses \cite{delhom2024entanglement}.
Dynamics beyond the PV model, such as secondary states or superpositions of primary states, may then arise as transient configurations during collisions involving $\geq3$ vortices.
Formally, the linear theory used in this work is valid when the intervortex distance goes to zero, whereas the PV model holds as this quantity tends to infinity.
Developing models that connect these two distinct regimes could therefore improve modelling of vortex dynamics across a wide scope of physical scenarios.
% A key feature of 2D turbulence is the inverse energy cascade, that is, energy tends to flow from small to large scales (the opposite of the usual cascade in 3D turbulence) manifesting as the formation of large vortex structures in the large $N$ limit. This tendency appears already for $N=3$ \cite{novikov1975dynamics}.

\textbf{Acknowledgements}. I am grateful to Zoran Hadzibabic for a question and ensuing discussion during the QSimFP Annual Workshop 2022, which prompted me to analyse close-range vortex states more closely.
I would also like to thank Ashton Bradley for insightful feedback on the manuscript.
This work was supported by the Science and Technology Facilities Council through the UKRI Quantum Technologies for Fundamental Physics Programme (Grant No. ST/T005858/1) and the Engineering and Physical Sciences Research Council through the Stephen Hawking Postdoctoral Fellowship (Grant No. EP/Z536660/1).

% \section*{}
% \bibliography{biblio}
\bibliography{version1.bbl}

\begin{thebibliography}{59}%
\makeatletter
\providecommand \@ifxundefined [1]{%
 \@ifx{#1\undefined}
}%
\providecommand \@ifnum [1]{%
 \ifnum #1\expandafter \@firstoftwo
 \else \expandafter \@secondoftwo
 \fi
}%
\providecommand \@ifx [1]{%
 \ifx #1\expandafter \@firstoftwo
 \else \expandafter \@secondoftwo
 \fi
}%
\providecommand \natexlab [1]{#1}%
\providecommand \enquote  [1]{``#1''}%
\providecommand \bibnamefont  [1]{#1}%
\providecommand \bibfnamefont [1]{#1}%
\providecommand \citenamefont [1]{#1}%
\providecommand \href@noop [0]{\@secondoftwo}%
\providecommand \href [0]{\begingroup \@sanitize@url \@href}%
\providecommand \@href[1]{\@@startlink{#1}\@@href}%
\providecommand \@@href[1]{\endgroup#1\@@endlink}%
\providecommand \@sanitize@url [0]{\catcode `\\12\catcode `\$12\catcode
  `\&12\catcode `\#12\catcode `\^12\catcode `\_12\catcode `\%12\relax}%
\providecommand \@@startlink[1]{}%
\providecommand \@@endlink[0]{}%
\providecommand \url  [0]{\begingroup\@sanitize@url \@url }%
\providecommand \@url [1]{\endgroup\@href {#1}{\urlprefix }}%
\providecommand \urlprefix  [0]{URL }%
\providecommand \Eprint [0]{\href }%
\providecommand \doibase [0]{https://doi.org/}%
\providecommand \selectlanguage [0]{\@gobble}%
\providecommand \bibinfo  [0]{\@secondoftwo}%
\providecommand \bibfield  [0]{\@secondoftwo}%
\providecommand \translation [1]{[#1]}%
\providecommand \BibitemOpen [0]{}%
\providecommand \bibitemStop [0]{}%
\providecommand \bibitemNoStop [0]{.\EOS\space}%
\providecommand \EOS [0]{\spacefactor3000\relax}%
\providecommand \BibitemShut  [1]{\csname bibitem#1\endcsname}%
\let\auto@bib@innerbib\@empty
%</preamble>
\bibitem [{\citenamefont {McWilliams}(1985)}]{mcwilliams1985submesoscale}%
  \BibitemOpen
  \bibfield  {author} {\bibinfo {author} {\bibfnamefont {J.~C.}\ \bibnamefont
  {McWilliams}},\ }\bibfield  {title} {\bibinfo {title} {Submesoscale, coherent
  vortices in the ocean},\ }\href@noop {} {\bibfield  {journal} {\bibinfo
  {journal} {Rev. Geophys.}\ }\textbf {\bibinfo {volume} {23}},\ \bibinfo
  {pages} {165} (\bibinfo {year} {1985})}\BibitemShut {NoStop}%
\bibitem [{\citenamefont {Kuo}(1966)}]{kuo1966dynamics}%
  \BibitemOpen
  \bibfield  {author} {\bibinfo {author} {\bibfnamefont {H.~L.}\ \bibnamefont
  {Kuo}},\ }\bibfield  {title} {\bibinfo {title} {On the dynamics of convective
  atmospheric vortices},\ }\href@noop {} {\bibfield  {journal} {\bibinfo
  {journal} {J. Atmos. Sci.}\ }\textbf {\bibinfo {volume} {23}},\ \bibinfo
  {pages} {25} (\bibinfo {year} {1966})}\BibitemShut {NoStop}%
\bibitem [{\citenamefont {Grier}(2003)}]{grier2003revolution}%
  \BibitemOpen
  \bibfield  {author} {\bibinfo {author} {\bibfnamefont {D.~G.}\ \bibnamefont
  {Grier}},\ }\bibfield  {title} {\bibinfo {title} {A revolution in optical
  manipulation},\ }\href@noop {} {\bibfield  {journal} {\bibinfo  {journal}
  {Nature}\ }\textbf {\bibinfo {volume} {424}},\ \bibinfo {pages} {810}
  (\bibinfo {year} {2003})}\BibitemShut {NoStop}%
\bibitem [{\citenamefont {Blatter}\ \emph {et~al.}(1994)\citenamefont
  {Blatter}, \citenamefont {Feigel'man}, \citenamefont {Geshkenbein},
  \citenamefont {Larkin},\ and\ \citenamefont {Vinokur}}]{blatter1994vortices}%
  \BibitemOpen
  \bibfield  {author} {\bibinfo {author} {\bibfnamefont {G.}~\bibnamefont
  {Blatter}}, \bibinfo {author} {\bibfnamefont {M.~V.}\ \bibnamefont
  {Feigel'man}}, \bibinfo {author} {\bibfnamefont {V.~B.}\ \bibnamefont
  {Geshkenbein}}, \bibinfo {author} {\bibfnamefont {A.~I.}\ \bibnamefont
  {Larkin}},\ and\ \bibinfo {author} {\bibfnamefont {V.~M.}\ \bibnamefont
  {Vinokur}},\ }\bibfield  {title} {\bibinfo {title} {Vortices in
  high-temperature superconductors},\ }\href@noop {} {\bibfield  {journal}
  {\bibinfo  {journal} {Rev. Mod. Phys.}\ }\textbf {\bibinfo {volume} {66}},\
  \bibinfo {pages} {1125} (\bibinfo {year} {1994})}\BibitemShut {NoStop}%
\bibitem [{\citenamefont {Wlaz{\l}owski}\ \emph {et~al.}(2016)\citenamefont
  {Wlaz{\l}owski}, \citenamefont {Sekizawa}, \citenamefont {Magierski},
  \citenamefont {Bulgac},\ and\ \citenamefont {Forbes}}]{wlazlowski2016vortex}%
  \BibitemOpen
  \bibfield  {author} {\bibinfo {author} {\bibfnamefont {G.}~\bibnamefont
  {Wlaz{\l}owski}}, \bibinfo {author} {\bibfnamefont {K.}~\bibnamefont
  {Sekizawa}}, \bibinfo {author} {\bibfnamefont {P.}~\bibnamefont {Magierski}},
  \bibinfo {author} {\bibfnamefont {A.}~\bibnamefont {Bulgac}},\ and\ \bibinfo
  {author} {\bibfnamefont {M.~M.}\ \bibnamefont {Forbes}},\ }\bibfield  {title}
  {\bibinfo {title} {Vortex pinning and dynamics in the neutron star crust},\
  }\href@noop {} {\bibfield  {journal} {\bibinfo  {journal} {Phys. Rev. Lett.}\
  }\textbf {\bibinfo {volume} {117}},\ \bibinfo {pages} {232701} (\bibinfo
  {year} {2016})}\BibitemShut {NoStop}%
\bibitem [{\citenamefont {Fetter}\ and\ \citenamefont
  {Svidzinsky}(2001)}]{fetter2001vortices}%
  \BibitemOpen
  \bibfield  {author} {\bibinfo {author} {\bibfnamefont {A.~L.}\ \bibnamefont
  {Fetter}}\ and\ \bibinfo {author} {\bibfnamefont {A.~A.}\ \bibnamefont
  {Svidzinsky}},\ }\bibfield  {title} {\bibinfo {title} {Vortices in a trapped
  dilute {Bose-Einstein} condensate},\ }\href@noop {} {\bibfield  {journal}
  {\bibinfo  {journal} {J. Phys. Condens. Matter}\ }\textbf {\bibinfo {volume}
  {13}},\ \bibinfo {pages} {R135} (\bibinfo {year} {2001})}\BibitemShut
  {NoStop}%
\bibitem [{\citenamefont {Barenghi}\ \emph {et~al.}(2001)\citenamefont
  {Barenghi}, \citenamefont {Donnelly},\ and\ \citenamefont
  {Vinen}}]{barenghi2001quantized}%
  \BibitemOpen
  \bibfield  {author} {\bibinfo {author} {\bibfnamefont {C.~F.}\ \bibnamefont
  {Barenghi}}, \bibinfo {author} {\bibfnamefont {R.~J.}\ \bibnamefont
  {Donnelly}},\ and\ \bibinfo {author} {\bibfnamefont {W.~F.}\ \bibnamefont
  {Vinen}},\ }\href@noop {} {\emph {\bibinfo {title} {Quantized vortex dynamics
  and superfluid turbulence}}},\ Vol.\ \bibinfo {volume} {571}\ (\bibinfo
  {publisher} {Springer Science \& Business Media},\ \bibinfo {year}
  {2001})\BibitemShut {NoStop}%
\bibitem [{\citenamefont {Carusotto}\ and\ \citenamefont
  {Ciuti}(2013)}]{carusotto2013quantum}%
  \BibitemOpen
  \bibfield  {author} {\bibinfo {author} {\bibfnamefont {I.}~\bibnamefont
  {Carusotto}}\ and\ \bibinfo {author} {\bibfnamefont {C.}~\bibnamefont
  {Ciuti}},\ }\bibfield  {title} {\bibinfo {title} {Quantum fluids of light},\
  }\href@noop {} {\bibfield  {journal} {\bibinfo  {journal} {Rev. Mod. Phys.}\
  }\textbf {\bibinfo {volume} {85}},\ \bibinfo {pages} {299} (\bibinfo {year}
  {2013})}\BibitemShut {NoStop}%
\bibitem [{\citenamefont {Ryzhov}\ \emph {et~al.}(2017)\citenamefont {Ryzhov},
  \citenamefont {Tareyeva}, \citenamefont {Fomin},\ and\ \citenamefont
  {Tsiok}}]{ryzhov2017berezinskii}%
  \BibitemOpen
  \bibfield  {author} {\bibinfo {author} {\bibfnamefont {V.~N.}\ \bibnamefont
  {Ryzhov}}, \bibinfo {author} {\bibfnamefont {E.~E.}\ \bibnamefont
  {Tareyeva}}, \bibinfo {author} {\bibfnamefont {Y.~D.}\ \bibnamefont
  {Fomin}},\ and\ \bibinfo {author} {\bibfnamefont {E.~N.}\ \bibnamefont
  {Tsiok}},\ }\bibfield  {title} {\bibinfo {title}
  {Berezinskii--kosterlitz--thouless transition and two-dimensional melting},\
  }\href@noop {} {\bibfield  {journal} {\bibinfo  {journal} {Physics-Uspekhi}\
  }\textbf {\bibinfo {volume} {60}},\ \bibinfo {pages} {857} (\bibinfo {year}
  {2017})}\BibitemShut {NoStop}%
\bibitem [{\citenamefont {Neely}\ \emph {et~al.}(2013)\citenamefont {Neely},
  \citenamefont {Bradley}, \citenamefont {Samson}, \citenamefont {Rooney},
  \citenamefont {Wright}, \citenamefont {Law}, \citenamefont
  {Carretero-Gonz{\'a}lez}, \citenamefont {Kevrekidis}, \citenamefont {Davis},\
  and\ \citenamefont {Anderson}}]{neely2013characteristics}%
  \BibitemOpen
  \bibfield  {author} {\bibinfo {author} {\bibfnamefont {T.~W.}\ \bibnamefont
  {Neely}}, \bibinfo {author} {\bibfnamefont {A.~S.}\ \bibnamefont {Bradley}},
  \bibinfo {author} {\bibfnamefont {E.~C.}\ \bibnamefont {Samson}}, \bibinfo
  {author} {\bibfnamefont {S.~J.}\ \bibnamefont {Rooney}}, \bibinfo {author}
  {\bibfnamefont {E.~M.}\ \bibnamefont {Wright}}, \bibinfo {author}
  {\bibfnamefont {K.~J.~H.}\ \bibnamefont {Law}}, \bibinfo {author}
  {\bibfnamefont {R.}~\bibnamefont {Carretero-Gonz{\'a}lez}}, \bibinfo {author}
  {\bibfnamefont {P.~G.}\ \bibnamefont {Kevrekidis}}, \bibinfo {author}
  {\bibfnamefont {M.~J.}\ \bibnamefont {Davis}},\ and\ \bibinfo {author}
  {\bibfnamefont {B.~P.}\ \bibnamefont {Anderson}},\ }\bibfield  {title}
  {\bibinfo {title} {Characteristics of two-dimensional quantum turbulence in a
  compressible superfluid},\ }\href@noop {} {\bibfield  {journal} {\bibinfo
  {journal} {Phys. Rev. Lett.}\ }\textbf {\bibinfo {volume} {111}},\ \bibinfo
  {pages} {235301} (\bibinfo {year} {2013})}\BibitemShut {NoStop}%
\bibitem [{\citenamefont {Kozik}\ and\ \citenamefont
  {Svistunov}(2009)}]{kozik2009theory}%
  \BibitemOpen
  \bibfield  {author} {\bibinfo {author} {\bibfnamefont {E.~V.}\ \bibnamefont
  {Kozik}}\ and\ \bibinfo {author} {\bibfnamefont {B.~V.}\ \bibnamefont
  {Svistunov}},\ }\bibfield  {title} {\bibinfo {title} {Theory of decay of
  superfluid turbulence in the low-temperature limit},\ }\href@noop {}
  {\bibfield  {journal} {\bibinfo  {journal} {J. Low. Temp. Phys.}\ }\textbf
  {\bibinfo {volume} {156}},\ \bibinfo {pages} {215} (\bibinfo {year}
  {2009})}\BibitemShut {NoStop}%
\bibitem [{\citenamefont {Groszek}\ \emph {et~al.}(2018)\citenamefont
  {Groszek}, \citenamefont {Paganin}, \citenamefont {Helmerson},\ and\
  \citenamefont {Simula}}]{groszek2018motion}%
  \BibitemOpen
  \bibfield  {author} {\bibinfo {author} {\bibfnamefont {A.~J.}\ \bibnamefont
  {Groszek}}, \bibinfo {author} {\bibfnamefont {D.~M.}\ \bibnamefont
  {Paganin}}, \bibinfo {author} {\bibfnamefont {K.}~\bibnamefont {Helmerson}},\
  and\ \bibinfo {author} {\bibfnamefont {T.~P.}\ \bibnamefont {Simula}},\
  }\bibfield  {title} {\bibinfo {title} {Motion of vortices in inhomogeneous
  {Bose-Einstein} condensates},\ }\href@noop {} {\bibfield  {journal} {\bibinfo
   {journal} {Phys. Rev. A}\ }\textbf {\bibinfo {volume} {97}},\ \bibinfo
  {pages} {023617} (\bibinfo {year} {2018})}\BibitemShut {NoStop}%
\bibitem [{\citenamefont {Carnevale}\ \emph {et~al.}(1991)\citenamefont
  {Carnevale}, \citenamefont {McWilliams}, \citenamefont {Weiss},\ and\
  \citenamefont {Young}}]{carnevale1991evolution}%
  \BibitemOpen
  \bibfield  {author} {\bibinfo {author} {\bibfnamefont {G.~F.}\ \bibnamefont
  {Carnevale}}, \bibinfo {author} {\bibfnamefont {Y.}~\bibnamefont
  {McWilliams}, \bibfnamefont {J.~C.and~Pomeau}}, \bibinfo {author}
  {\bibfnamefont {J.~B.}\ \bibnamefont {Weiss}},\ and\ \bibinfo {author}
  {\bibfnamefont {W.~R.}\ \bibnamefont {Young}},\ }\bibfield  {title} {\bibinfo
  {title} {Evolution of vortex statistics in two-dimensional turbulence},\
  }\href@noop {} {\bibfield  {journal} {\bibinfo  {journal} {Phys. Rev. Lett}\
  }\textbf {\bibinfo {volume} {66}},\ \bibinfo {pages} {2735} (\bibinfo {year}
  {1991})}\BibitemShut {NoStop}%
\bibitem [{\citenamefont {Shin}\ \emph {et~al.}(2004)\citenamefont {Shin},
  \citenamefont {Saba}, \citenamefont {Vengalattore}, \citenamefont {Pasquini},
  \citenamefont {Sanner}, \citenamefont {Leanhardt}, \citenamefont {Prentiss},
  \citenamefont {Pritchard},\ and\ \citenamefont
  {Ketterle}}]{shin2004dynamical}%
  \BibitemOpen
  \bibfield  {author} {\bibinfo {author} {\bibfnamefont {Y.}~\bibnamefont
  {Shin}}, \bibinfo {author} {\bibfnamefont {M.}~\bibnamefont {Saba}}, \bibinfo
  {author} {\bibfnamefont {M.}~\bibnamefont {Vengalattore}}, \bibinfo {author}
  {\bibfnamefont {T.~A.}\ \bibnamefont {Pasquini}}, \bibinfo {author}
  {\bibfnamefont {C.}~\bibnamefont {Sanner}}, \bibinfo {author} {\bibfnamefont
  {A.~E.}\ \bibnamefont {Leanhardt}}, \bibinfo {author} {\bibfnamefont
  {M.}~\bibnamefont {Prentiss}}, \bibinfo {author} {\bibfnamefont {D.~E.}\
  \bibnamefont {Pritchard}},\ and\ \bibinfo {author} {\bibfnamefont
  {W.}~\bibnamefont {Ketterle}},\ }\bibfield  {title} {\bibinfo {title}
  {Dynamical instability of a doubly quantized vortex in a {Bose-Einstein}
  condensate},\ }\href@noop {} {\bibfield  {journal} {\bibinfo  {journal}
  {Phys. Rev. Lett.}\ }\textbf {\bibinfo {volume} {93}},\ \bibinfo {pages}
  {160406} (\bibinfo {year} {2004})}\BibitemShut {NoStop}%
\bibitem [{\citenamefont {Kawaguchi}\ and\ \citenamefont
  {Ohmi}(2004)}]{kawaguchi2004splitting}%
  \BibitemOpen
  \bibfield  {author} {\bibinfo {author} {\bibfnamefont {Y.}~\bibnamefont
  {Kawaguchi}}\ and\ \bibinfo {author} {\bibfnamefont {T.}~\bibnamefont
  {Ohmi}},\ }\bibfield  {title} {\bibinfo {title} {Splitting instability of a
  multiply charged vortex in a {Bose-Einstein} condensate},\ }\href@noop {}
  {\bibfield  {journal} {\bibinfo  {journal} {Phys. Rev. A}\ }\textbf {\bibinfo
  {volume} {70}},\ \bibinfo {pages} {043610} (\bibinfo {year}
  {2004})}\BibitemShut {NoStop}%
\bibitem [{\citenamefont {Kumakura}\ \emph {et~al.}(2006)\citenamefont
  {Kumakura}, \citenamefont {Hirotani}, \citenamefont {Okano}, \citenamefont
  {Takahashi},\ and\ \citenamefont {Yabuzaki}}]{kumakura2006topological}%
  \BibitemOpen
  \bibfield  {author} {\bibinfo {author} {\bibfnamefont {M.}~\bibnamefont
  {Kumakura}}, \bibinfo {author} {\bibfnamefont {T.}~\bibnamefont {Hirotani}},
  \bibinfo {author} {\bibfnamefont {M.}~\bibnamefont {Okano}}, \bibinfo
  {author} {\bibfnamefont {Y.}~\bibnamefont {Takahashi}},\ and\ \bibinfo
  {author} {\bibfnamefont {T.}~\bibnamefont {Yabuzaki}},\ }\bibfield  {title}
  {\bibinfo {title} {Topological formation of a multiply charged vortex in the
  {Rb Bose-Einstein} condensate: Effectiveness of the gravity compensation},\
  }\href@noop {} {\bibfield  {journal} {\bibinfo  {journal} {Phys. Rev. A}\
  }\textbf {\bibinfo {volume} {73}},\ \bibinfo {pages} {063605} (\bibinfo
  {year} {2006})}\BibitemShut {NoStop}%
\bibitem [{\citenamefont {Okano}\ \emph {et~al.}(2007)\citenamefont {Okano},
  \citenamefont {Yasuda}, \citenamefont {Kasa}, \citenamefont {Kumakura},\ and\
  \citenamefont {Takahashi}}]{okano2007splitting}%
  \BibitemOpen
  \bibfield  {author} {\bibinfo {author} {\bibfnamefont {M.}~\bibnamefont
  {Okano}}, \bibinfo {author} {\bibfnamefont {H.}~\bibnamefont {Yasuda}},
  \bibinfo {author} {\bibfnamefont {K.}~\bibnamefont {Kasa}}, \bibinfo {author}
  {\bibfnamefont {M.}~\bibnamefont {Kumakura}},\ and\ \bibinfo {author}
  {\bibfnamefont {Y.}~\bibnamefont {Takahashi}},\ }\bibfield  {title} {\bibinfo
  {title} {Splitting of a quadruply quantized vortex in the {Rb Bose-Einstein}
  condensate},\ }\href@noop {} {\bibfield  {journal} {\bibinfo  {journal} {J.
  Low Temp. Phys.}\ }\textbf {\bibinfo {volume} {148}},\ \bibinfo {pages} {447}
  (\bibinfo {year} {2007})}\BibitemShut {NoStop}%
\bibitem [{\citenamefont {Isoshima}\ \emph {et~al.}(2007)\citenamefont
  {Isoshima}, \citenamefont {Okano}, \citenamefont {Yasuda}, \citenamefont
  {Kasa}, \citenamefont {Huhtam{\"a}ki}, \citenamefont {Kumakura},\ and\
  \citenamefont {Takahashi}}]{isoshima2007spontaneous}%
  \BibitemOpen
  \bibfield  {author} {\bibinfo {author} {\bibfnamefont {T.}~\bibnamefont
  {Isoshima}}, \bibinfo {author} {\bibfnamefont {M.}~\bibnamefont {Okano}},
  \bibinfo {author} {\bibfnamefont {H.}~\bibnamefont {Yasuda}}, \bibinfo
  {author} {\bibfnamefont {K.}~\bibnamefont {Kasa}}, \bibinfo {author}
  {\bibfnamefont {J.~A.~M.}\ \bibnamefont {Huhtam{\"a}ki}}, \bibinfo {author}
  {\bibfnamefont {M.}~\bibnamefont {Kumakura}},\ and\ \bibinfo {author}
  {\bibfnamefont {Y.}~\bibnamefont {Takahashi}},\ }\bibfield  {title} {\bibinfo
  {title} {Spontaneous splitting of a quadruply charged vortex},\ }\href@noop
  {} {\bibfield  {journal} {\bibinfo  {journal} {Phys. Rev. Lett.}\ }\textbf
  {\bibinfo {volume} {99}},\ \bibinfo {pages} {200403} (\bibinfo {year}
  {2007})}\BibitemShut {NoStop}%
\bibitem [{\citenamefont {Kuwamoto}\ \emph {et~al.}(2010)\citenamefont
  {Kuwamoto}, \citenamefont {Usuda}, \citenamefont {Tojo},\ and\ \citenamefont
  {Hirano}}]{kuwamoto2010dynamics}%
  \BibitemOpen
  \bibfield  {author} {\bibinfo {author} {\bibfnamefont {T.}~\bibnamefont
  {Kuwamoto}}, \bibinfo {author} {\bibfnamefont {H.}~\bibnamefont {Usuda}},
  \bibinfo {author} {\bibfnamefont {S.}~\bibnamefont {Tojo}},\ and\ \bibinfo
  {author} {\bibfnamefont {T.}~\bibnamefont {Hirano}},\ }\bibfield  {title}
  {\bibinfo {title} {Dynamics of quadruply quantized vortices in {87Rb
  Bose--Einstein} condensates confined in magnetic and optical traps},\
  }\href@noop {} {\bibfield  {journal} {\bibinfo  {journal} {J, Phys. Soc.
  Japan}\ }\textbf {\bibinfo {volume} {79}},\ \bibinfo {pages} {034004}
  (\bibinfo {year} {2010})}\BibitemShut {NoStop}%
\bibitem [{\citenamefont {Shibayama}\ \emph {et~al.}(2016)\citenamefont
  {Shibayama}, \citenamefont {Tsukada}, \citenamefont {Yoshihara},\ and\
  \citenamefont {Kuwamoto}}]{shibayama2016density}%
  \BibitemOpen
  \bibfield  {author} {\bibinfo {author} {\bibfnamefont {H.}~\bibnamefont
  {Shibayama}}, \bibinfo {author} {\bibfnamefont {A.}~\bibnamefont {Tsukada}},
  \bibinfo {author} {\bibfnamefont {T.}~\bibnamefont {Yoshihara}},\ and\
  \bibinfo {author} {\bibfnamefont {T.}~\bibnamefont {Kuwamoto}},\ }\bibfield
  {title} {\bibinfo {title} {Density dependence of charge-4 vortex splitting in
  {Bose--Einstein} condensates},\ }\href@noop {} {\bibfield  {journal}
  {\bibinfo  {journal} {J. Phys. Soc. Japan}\ }\textbf {\bibinfo {volume}
  {85}},\ \bibinfo {pages} {054401} (\bibinfo {year} {2016})}\BibitemShut
  {NoStop}%
\bibitem [{\citenamefont {Pu}\ \emph {et~al.}(1999)\citenamefont {Pu},
  \citenamefont {Law}, \citenamefont {Eberly},\ and\ \citenamefont
  {Bigelow}}]{pu1999coherent}%
  \BibitemOpen
  \bibfield  {author} {\bibinfo {author} {\bibfnamefont {H.}~\bibnamefont
  {Pu}}, \bibinfo {author} {\bibfnamefont {C.~K.}\ \bibnamefont {Law}},
  \bibinfo {author} {\bibfnamefont {J.~H.}\ \bibnamefont {Eberly}},\ and\
  \bibinfo {author} {\bibfnamefont {N.~P.}\ \bibnamefont {Bigelow}},\
  }\bibfield  {title} {\bibinfo {title} {Coherent disintegration and stability
  of vortices in trapped {Bose} condensates},\ }\href@noop {} {\bibfield
  {journal} {\bibinfo  {journal} {Phys. Rev. A}\ }\textbf {\bibinfo {volume}
  {59}},\ \bibinfo {pages} {1533} (\bibinfo {year} {1999})}\BibitemShut
  {NoStop}%
\bibitem [{\citenamefont {Nilsen}\ and\ \citenamefont
  {Lundh}(2008)}]{nilsen2008splitting}%
  \BibitemOpen
  \bibfield  {author} {\bibinfo {author} {\bibfnamefont {H.~M.}\ \bibnamefont
  {Nilsen}}\ and\ \bibinfo {author} {\bibfnamefont {E.}~\bibnamefont {Lundh}},\
  }\bibfield  {title} {\bibinfo {title} {Splitting dynamics of doubly quantized
  vortices in {Bose-Einstein} condensates},\ }\href@noop {} {\bibfield
  {journal} {\bibinfo  {journal} {Phys. Rev. A}\ }\textbf {\bibinfo {volume}
  {77}},\ \bibinfo {pages} {013604} (\bibinfo {year} {2008})}\BibitemShut
  {NoStop}%
\bibitem [{\citenamefont {Zhu}\ and\ \citenamefont
  {Pan}(2021)}]{zhu2021splitting}%
  \BibitemOpen
  \bibfield  {author} {\bibinfo {author} {\bibfnamefont {Q.}~\bibnamefont
  {Zhu}}\ and\ \bibinfo {author} {\bibfnamefont {L.}~\bibnamefont {Pan}},\
  }\bibfield  {title} {\bibinfo {title} {Splitting of a multiply quantized
  vortex for a {Bose-Einstein} condensate in an optical lattice},\ }\href@noop
  {} {\bibfield  {journal} {\bibinfo  {journal} {J. Low. Temp. Phys.}\ }\textbf
  {\bibinfo {volume} {203}},\ \bibinfo {pages} {392} (\bibinfo {year}
  {2021})}\BibitemShut {NoStop}%
\bibitem [{\citenamefont {Patrick}\ \emph
  {et~al.}(2022{\natexlab{a}})\citenamefont {Patrick}, \citenamefont
  {Geelmuyden}, \citenamefont {Erne}, \citenamefont {Barenghi},\ and\
  \citenamefont {Weinfurtner}}]{patrick2022origin}%
  \BibitemOpen
  \bibfield  {author} {\bibinfo {author} {\bibfnamefont {S.}~\bibnamefont
  {Patrick}}, \bibinfo {author} {\bibfnamefont {A.}~\bibnamefont {Geelmuyden}},
  \bibinfo {author} {\bibfnamefont {S.}~\bibnamefont {Erne}}, \bibinfo {author}
  {\bibfnamefont {C.~F.}\ \bibnamefont {Barenghi}},\ and\ \bibinfo {author}
  {\bibfnamefont {S.}~\bibnamefont {Weinfurtner}},\ }\bibfield  {title}
  {\bibinfo {title} {Origin and evolution of the multiply quantized vortex
  instability},\ }\href@noop {} {\bibfield  {journal} {\bibinfo  {journal}
  {Phys. Rev. Research}\ }\textbf {\bibinfo {volume} {4}},\ \bibinfo {pages}
  {043104} (\bibinfo {year} {2022}{\natexlab{a}})}\BibitemShut {NoStop}%
\bibitem [{\citenamefont {Lan}\ \emph {et~al.}(2023{\natexlab{a}})\citenamefont
  {Lan}, \citenamefont {Li}, \citenamefont {Mo}, \citenamefont {Tian},
  \citenamefont {Yan}, \citenamefont {Yang},\ and\ \citenamefont
  {Zhang}}]{lan2023splitting}%
  \BibitemOpen
  \bibfield  {author} {\bibinfo {author} {\bibfnamefont {S.}~\bibnamefont
  {Lan}}, \bibinfo {author} {\bibfnamefont {X.}~\bibnamefont {Li}}, \bibinfo
  {author} {\bibfnamefont {J.}~\bibnamefont {Mo}}, \bibinfo {author}
  {\bibfnamefont {Y.}~\bibnamefont {Tian}}, \bibinfo {author} {\bibfnamefont
  {Y.}~\bibnamefont {Yan}}, \bibinfo {author} {\bibfnamefont {P.}~\bibnamefont
  {Yang}},\ and\ \bibinfo {author} {\bibfnamefont {H.}~\bibnamefont {Zhang}},\
  }\bibfield  {title} {\bibinfo {title} {Splitting of doubly quantized vortices
  in holographic superfluid of finite temperature},\ }\href@noop {} {\bibfield
  {journal} {\bibinfo  {journal} {J. High Energy Phys.}\ }\textbf {\bibinfo
  {volume} {2023}}\bibinfo  {number} { (5)},\ \bibinfo {pages} {1}}\BibitemShut
  {NoStop}%
\bibitem [{\citenamefont {Lan}\ \emph {et~al.}(2023{\natexlab{b}})\citenamefont
  {Lan}, \citenamefont {Li}, \citenamefont {Tian}, \citenamefont {Yang},\ and\
  \citenamefont {Zhang}}]{lan2023heating}%
  \BibitemOpen
\bibfield  {number} {  }\bibfield  {author} {\bibinfo {author} {\bibfnamefont
  {S.}~\bibnamefont {Lan}}, \bibinfo {author} {\bibfnamefont {X.}~\bibnamefont
  {Li}}, \bibinfo {author} {\bibfnamefont {Y.}~\bibnamefont {Tian}}, \bibinfo
  {author} {\bibfnamefont {P.}~\bibnamefont {Yang}},\ and\ \bibinfo {author}
  {\bibfnamefont {H.}~\bibnamefont {Zhang}},\ }\bibfield  {title} {\bibinfo
  {title} {Heating up quadruply quantized vortices: Splitting patterns and
  dynamical transitions},\ }\href@noop {} {\bibfield  {journal} {\bibinfo
  {journal} {Phys. Rev. Lett.}\ }\textbf {\bibinfo {volume} {131}},\ \bibinfo
  {pages} {221602} (\bibinfo {year} {2023}{\natexlab{b}})}\BibitemShut
  {NoStop}%
\bibitem [{\citenamefont {Adhikari}(2019)}]{adhikari2019stable}%
  \BibitemOpen
  \bibfield  {author} {\bibinfo {author} {\bibfnamefont {S.~K.}\ \bibnamefont
  {Adhikari}},\ }\bibfield  {title} {\bibinfo {title} {Stable controllable
  giant vortex in a trapped {Bose--Einstein} condensate},\ }\href@noop {}
  {\bibfield  {journal} {\bibinfo  {journal} {Laser Phys. Lett.}\ }\textbf
  {\bibinfo {volume} {16}},\ \bibinfo {pages} {085501} (\bibinfo {year}
  {2019})}\BibitemShut {NoStop}%
\bibitem [{\citenamefont {Srinivasan}(2006)}]{srinivasan2006vortices}%
  \BibitemOpen
  \bibfield  {author} {\bibinfo {author} {\bibfnamefont {R.}~\bibnamefont
  {Srinivasan}},\ }\bibfield  {title} {\bibinfo {title} {Vortices in
  {Bose-Einstein condensates}: A review of the experimental results},\
  }\href@noop {} {\bibfield  {journal} {\bibinfo  {journal} {Pramana}\ }\textbf
  {\bibinfo {volume} {66}},\ \bibinfo {pages} {3} (\bibinfo {year}
  {2006})}\BibitemShut {NoStop}%
\bibitem [{\citenamefont {Alperin}\ and\ \citenamefont
  {Berloff}(2021)}]{alperin2021multiply}%
  \BibitemOpen
  \bibfield  {author} {\bibinfo {author} {\bibfnamefont {S.~N.}\ \bibnamefont
  {Alperin}}\ and\ \bibinfo {author} {\bibfnamefont {N.~G.}\ \bibnamefont
  {Berloff}},\ }\bibfield  {title} {\bibinfo {title} {Multiply charged vortex
  states of polariton condensates},\ }\href@noop {} {\bibfield  {journal}
  {\bibinfo  {journal} {Optica}\ }\textbf {\bibinfo {volume} {8}},\ \bibinfo
  {pages} {301} (\bibinfo {year} {2021})}\BibitemShut {NoStop}%
\bibitem [{\citenamefont {Ruffenach}\ \emph {et~al.}(2023)\citenamefont
  {Ruffenach}, \citenamefont {Galantucci},\ and\ \citenamefont
  {Barenghi}}]{ruffenach2023superfluid}%
  \BibitemOpen
  \bibfield  {author} {\bibinfo {author} {\bibfnamefont {W.}~\bibnamefont
  {Ruffenach}}, \bibinfo {author} {\bibfnamefont {L.}~\bibnamefont
  {Galantucci}},\ and\ \bibinfo {author} {\bibfnamefont {C.~F.}\ \bibnamefont
  {Barenghi}},\ }\bibfield  {title} {\bibinfo {title} {Superfluid drain
  vortex},\ }\href@noop {} {\bibfield  {journal} {\bibinfo  {journal} {J. Low
  Temp. Phys.}\ }\textbf {\bibinfo {volume} {212}},\ \bibinfo {pages} {375}
  (\bibinfo {year} {2023})}\BibitemShut {NoStop}%
\bibitem [{\citenamefont {Delhom}\ \emph {et~al.}(2024)\citenamefont {Delhom},
  \citenamefont {Guerrero}, \citenamefont {Calizaya~Cabrera}, \citenamefont
  {Falque}, \citenamefont {Bramati}, \citenamefont {Brady}, \citenamefont
  {Jacquet},\ and\ \citenamefont {Agullo}}]{delhom2024entanglement}%
  \BibitemOpen
  \bibfield  {author} {\bibinfo {author} {\bibfnamefont {A.}~\bibnamefont
  {Delhom}}, \bibinfo {author} {\bibfnamefont {K.}~\bibnamefont {Guerrero}},
  \bibinfo {author} {\bibfnamefont {P.}~\bibnamefont {Calizaya~Cabrera}},
  \bibinfo {author} {\bibfnamefont {K.}~\bibnamefont {Falque}}, \bibinfo
  {author} {\bibfnamefont {A.}~\bibnamefont {Bramati}}, \bibinfo {author}
  {\bibfnamefont {A.~J.}\ \bibnamefont {Brady}}, \bibinfo {author}
  {\bibfnamefont {M.~J.}\ \bibnamefont {Jacquet}},\ and\ \bibinfo {author}
  {\bibfnamefont {I.}~\bibnamefont {Agullo}},\ }\bibfield  {title} {\bibinfo
  {title} {Entanglement from superradiance and rotating quantum fluids of
  light},\ }\href@noop {} {\bibfield  {journal} {\bibinfo  {journal} {Phys.
  Rev. D}\ }\textbf {\bibinfo {volume} {109}},\ \bibinfo {pages} {105024}
  (\bibinfo {year} {2024})}\BibitemShut {NoStop}%
\bibitem [{\citenamefont {Patrick}(2024)}]{patrick2024quantum}%
  \BibitemOpen
  \bibfield  {author} {\bibinfo {author} {\bibfnamefont {S.}~\bibnamefont
  {Patrick}},\ }\bibfield  {title} {\bibinfo {title} {Quantum vortex stability
  in draining fluid flows},\ }\href@noop {} {\bibfield  {journal} {\bibinfo
  {journal} {Phys. Rev. A}\ }\textbf {\bibinfo {volume} {110}},\ \bibinfo
  {pages} {013327} (\bibinfo {year} {2024})}\BibitemShut {NoStop}%
\bibitem [{\citenamefont {{\v{S}}van{\v{c}}ara}\ \emph
  {et~al.}(2024)\citenamefont {{\v{S}}van{\v{c}}ara}, \citenamefont
  {Smaniotto}, \citenamefont {Solidoro}, \citenamefont {MacDonald},
  \citenamefont {Patrick}, \citenamefont {Gregory}, \citenamefont {Barenghi},\
  and\ \citenamefont {Weinfurtner}}]{svanvcara2024rotating}%
  \BibitemOpen
  \bibfield  {author} {\bibinfo {author} {\bibfnamefont {P.}~\bibnamefont
  {{\v{S}}van{\v{c}}ara}}, \bibinfo {author} {\bibfnamefont {P.}~\bibnamefont
  {Smaniotto}}, \bibinfo {author} {\bibfnamefont {L.}~\bibnamefont {Solidoro}},
  \bibinfo {author} {\bibfnamefont {J.~F.}\ \bibnamefont {MacDonald}}, \bibinfo
  {author} {\bibfnamefont {S.}~\bibnamefont {Patrick}}, \bibinfo {author}
  {\bibfnamefont {R.}~\bibnamefont {Gregory}}, \bibinfo {author} {\bibfnamefont
  {C.~F.}\ \bibnamefont {Barenghi}},\ and\ \bibinfo {author} {\bibfnamefont
  {S.}~\bibnamefont {Weinfurtner}},\ }\bibfield  {title} {\bibinfo {title}
  {Rotating curved spacetime signatures from a giant quantum vortex},\
  }\href@noop {} {\bibfield  {journal} {\bibinfo  {journal} {Nature}\ }\textbf
  {\bibinfo {volume} {628}},\ \bibinfo {pages} {66} (\bibinfo {year}
  {2024})}\BibitemShut {NoStop}%
\bibitem [{\citenamefont {Wang}\ \emph {et~al.}(2011)\citenamefont {Wang},
  \citenamefont {Song}, \citenamefont {Xiong},\ and\ \citenamefont
  {Liu}}]{wang2011quantized}%
  \BibitemOpen
  \bibfield  {author} {\bibinfo {author} {\bibfnamefont {D.}~\bibnamefont
  {Wang}}, \bibinfo {author} {\bibfnamefont {S.}~\bibnamefont {Song}}, \bibinfo
  {author} {\bibfnamefont {B.}~\bibnamefont {Xiong}},\ and\ \bibinfo {author}
  {\bibfnamefont {W.}~\bibnamefont {Liu}},\ }\bibfield  {title} {\bibinfo
  {title} {Quantized vortices in a rotating {Bose-Einstein} condensate with
  spatiotemporally modulated interaction},\ }\href@noop {} {\bibfield
  {journal} {\bibinfo  {journal} {Phys. Rev. A}\ }\textbf {\bibinfo {volume}
  {84}},\ \bibinfo {pages} {053607} (\bibinfo {year} {2011})}\BibitemShut
  {NoStop}%
\bibitem [{\citenamefont {Patrick}\ \emph {et~al.}(2023)\citenamefont
  {Patrick}, \citenamefont {Gupta}, \citenamefont {Gregory},\ and\
  \citenamefont {Barenghi}}]{patrick2023stability}%
  \BibitemOpen
  \bibfield  {author} {\bibinfo {author} {\bibfnamefont {S.}~\bibnamefont
  {Patrick}}, \bibinfo {author} {\bibfnamefont {A.}~\bibnamefont {Gupta}},
  \bibinfo {author} {\bibfnamefont {R.}~\bibnamefont {Gregory}},\ and\ \bibinfo
  {author} {\bibfnamefont {C.~F.}\ \bibnamefont {Barenghi}},\ }\bibfield
  {title} {\bibinfo {title} {Stability of quantized vortices in two-component
  condensates},\ }\href@noop {} {\bibfield  {journal} {\bibinfo  {journal}
  {Phys. Rev. Research}\ }\textbf {\bibinfo {volume} {5}},\ \bibinfo {pages}
  {033201} (\bibinfo {year} {2023})}\BibitemShut {NoStop}%
\bibitem [{\citenamefont {Berti}\ \emph {et~al.}(2023)\citenamefont {Berti},
  \citenamefont {Giacomelli},\ and\ \citenamefont
  {Carusotto}}]{berti2023superradiant}%
  \BibitemOpen
  \bibfield  {author} {\bibinfo {author} {\bibfnamefont {A.}~\bibnamefont
  {Berti}}, \bibinfo {author} {\bibfnamefont {L.}~\bibnamefont {Giacomelli}},\
  and\ \bibinfo {author} {\bibfnamefont {I.}~\bibnamefont {Carusotto}},\
  }\bibfield  {title} {\bibinfo {title} {Superradiant phononic emission from
  the analog spin ergoregion in a two-component {Bose--Einstein} condensate},\
  }\href@noop {} {\bibfield  {journal} {\bibinfo  {journal} {Comptes Rendus.
  Physique}\ }\textbf {\bibinfo {volume} {24}},\ \bibinfo {pages} {113}
  (\bibinfo {year} {2023})}\BibitemShut {NoStop}%
\bibitem [{\citenamefont {Kuopanportti}\ \emph {et~al.}(2015)\citenamefont
  {Kuopanportti}, \citenamefont {Orlova},\ and\ \citenamefont
  {Milo{\v{s}}evi{\'c}}}]{kuopanportti2015ground}%
  \BibitemOpen
  \bibfield  {author} {\bibinfo {author} {\bibfnamefont {P.}~\bibnamefont
  {Kuopanportti}}, \bibinfo {author} {\bibfnamefont {N.~V.}\ \bibnamefont
  {Orlova}},\ and\ \bibinfo {author} {\bibfnamefont {M.~V.}\ \bibnamefont
  {Milo{\v{s}}evi{\'c}}},\ }\bibfield  {title} {\bibinfo {title} {Ground-state
  multiquantum vortices in rotating two-species superfluids},\ }\href@noop {}
  {\bibfield  {journal} {\bibinfo  {journal} {Phys. Rev. A}\ }\textbf {\bibinfo
  {volume} {91}},\ \bibinfo {pages} {043605} (\bibinfo {year}
  {2015})}\BibitemShut {NoStop}%
\bibitem [{\citenamefont {An}\ and\ \citenamefont
  {Li}(2025)}]{an2025splitting}%
  \BibitemOpen
  \bibfield  {author} {\bibinfo {author} {\bibfnamefont {Y.}~\bibnamefont
  {An}}\ and\ \bibinfo {author} {\bibfnamefont {L.}~\bibnamefont {Li}},\
  }\bibfield  {title} {\bibinfo {title} {Splitting dynamics of quantized
  composite vortices in holographic miscible binary superfluids},\ }\href@noop
  {} {\bibfield  {journal} {\bibinfo  {journal} {arXiv preprint
  arXiv:2501.03561}\ } (\bibinfo {year} {2025})}\BibitemShut {NoStop}%
\bibitem [{\citenamefont {Cockburn}\ and\ \citenamefont
  {Proukakis}(2009)}]{cockburn2009stochastic}%
  \BibitemOpen
  \bibfield  {author} {\bibinfo {author} {\bibfnamefont {S.~P.}\ \bibnamefont
  {Cockburn}}\ and\ \bibinfo {author} {\bibfnamefont {N.~P.}\ \bibnamefont
  {Proukakis}},\ }\bibfield  {title} {\bibinfo {title} {The stochastic
  {Gross-Pitaevskii} equation and some applications},\ }\href@noop {}
  {\bibfield  {journal} {\bibinfo  {journal} {Laser Phys.}\ }\textbf {\bibinfo
  {volume} {19}},\ \bibinfo {pages} {558} (\bibinfo {year} {2009})}\BibitemShut
  {NoStop}%
\bibitem [{\citenamefont {Patrick}\ \emph
  {et~al.}(2022{\natexlab{b}})\citenamefont {Patrick}, \citenamefont
  {Geelmuyden}, \citenamefont {Erne}, \citenamefont {Barenghi},\ and\
  \citenamefont {Weinfurtner}}]{patrick2022quantum}%
  \BibitemOpen
  \bibfield  {author} {\bibinfo {author} {\bibfnamefont {S.}~\bibnamefont
  {Patrick}}, \bibinfo {author} {\bibfnamefont {A.}~\bibnamefont {Geelmuyden}},
  \bibinfo {author} {\bibfnamefont {S.}~\bibnamefont {Erne}}, \bibinfo {author}
  {\bibfnamefont {C.~F.}\ \bibnamefont {Barenghi}},\ and\ \bibinfo {author}
  {\bibfnamefont {S.}~\bibnamefont {Weinfurtner}},\ }\bibfield  {title}
  {\bibinfo {title} {Quantum vortex instability and black hole superradiance},\
  }\href@noop {} {\bibfield  {journal} {\bibinfo  {journal} {Phys. Rev.
  Research}\ }\textbf {\bibinfo {volume} {4}},\ \bibinfo {pages} {033117}
  (\bibinfo {year} {2022}{\natexlab{b}})}\BibitemShut {NoStop}%
\bibitem [{\citenamefont {Giacomelli}\ and\ \citenamefont
  {Carusotto}(2020)}]{giacomelli2020ergoregion}%
  \BibitemOpen
  \bibfield  {author} {\bibinfo {author} {\bibfnamefont {L.}~\bibnamefont
  {Giacomelli}}\ and\ \bibinfo {author} {\bibfnamefont {I.}~\bibnamefont
  {Carusotto}},\ }\bibfield  {title} {\bibinfo {title} {Ergoregion
  instabilities in rotating two-dimensional {Bose-Einstein} condensates:
  {Perspectives} on the stability of quantized vortices},\ }\href@noop {}
  {\bibfield  {journal} {\bibinfo  {journal} {Phys. Rev. Research}\ }\textbf
  {\bibinfo {volume} {2}},\ \bibinfo {pages} {033139} (\bibinfo {year}
  {2020})}\BibitemShut {NoStop}%
\bibitem [{Note1()}]{Note1}%
  \BibitemOpen
  \bibinfo {note} {Simulations were performed on a square grid $x\in [-L,L]$,
  with $L=r_B+5$ and $N_x=128$ grid points (similarly for $y$). The time step
  was $\Delta t=L^2/5N_x^2$. The initial vortex profile $\protect \sqrt
  {\protect \bar {n}}e^{i\ell \theta }$ was found by evolving Eq.~\protect
  \eqref {GPE} in imaginary time by factoring out the phase and simulating only
  the radial direction. Eigenmodes are obtained by diagonalising $\omega
  _\lambda |\psi _\lambda \rangle = \protect \mathcal {L}|\psi _\lambda \rangle
  $, approximating the radial derivatives with 5-point centred finite
  difference stencils everywhere except $r=0$ ($r=L$) where we apply a 5-point
  forward (backward) stencil and a regularity (Dirichlet) boundary
  condition.}\BibitemShut {Stop}%
\bibitem [{\citenamefont {Timmermans}\ \emph {et~al.}(1999)\citenamefont
  {Timmermans}, \citenamefont {Tommasini}, \citenamefont {Hussein},\ and\
  \citenamefont {Kerman}}]{timmermans1999feshbach}%
  \BibitemOpen
  \bibfield  {author} {\bibinfo {author} {\bibfnamefont {E.}~\bibnamefont
  {Timmermans}}, \bibinfo {author} {\bibfnamefont {P.}~\bibnamefont
  {Tommasini}}, \bibinfo {author} {\bibfnamefont {M.}~\bibnamefont {Hussein}},\
  and\ \bibinfo {author} {\bibfnamefont {A.}~\bibnamefont {Kerman}},\
  }\bibfield  {title} {\bibinfo {title} {Feshbach resonances in atomic
  {Bose--Einstein} condensates},\ }\href@noop {} {\bibfield  {journal}
  {\bibinfo  {journal} {Phys. Rep.}\ }\textbf {\bibinfo {volume} {315}},\
  \bibinfo {pages} {199} (\bibinfo {year} {1999})}\BibitemShut {NoStop}%
\bibitem [{\citenamefont {Caradoc-Davies}\ \emph {et~al.}(1999)\citenamefont
  {Caradoc-Davies}, \citenamefont {Ballagh},\ and\ \citenamefont
  {Burnett}}]{caradoc1999coherent}%
  \BibitemOpen
  \bibfield  {author} {\bibinfo {author} {\bibfnamefont {B.~M.}\ \bibnamefont
  {Caradoc-Davies}}, \bibinfo {author} {\bibfnamefont {R.~J.}\ \bibnamefont
  {Ballagh}},\ and\ \bibinfo {author} {\bibfnamefont {K.}~\bibnamefont
  {Burnett}},\ }\bibfield  {title} {\bibinfo {title} {Coherent dynamics of
  vortex formation in trapped {Bose-Einstein} condensates},\ }\href@noop {}
  {\bibfield  {journal} {\bibinfo  {journal} {Phys. Rev. Lett.}\ }\textbf
  {\bibinfo {volume} {83}},\ \bibinfo {pages} {895} (\bibinfo {year}
  {1999})}\BibitemShut {NoStop}%
\bibitem [{Note2()}]{Note2}%
  \BibitemOpen
  \bibinfo {note} {The seeding of a linear mode in the nonlinear equation
  \protect \eqref {GPE} introduces a correction to the vortex density due to
  backreaction effects. This induces a small shift in the chemical potential
  $\mu $ which is calculated by measuring the phase drift in $\Psi $ over the
  duration of the simulation. Using the corrected chemical potential, the
  normal modes of the vortex are recomputed. This procedure slightly shifts the
  $\omega _\lambda $, hence why $\zeta $ is quoted to 2 decimal
  places.}\BibitemShut {Stop}%
\bibitem [{\citenamefont {Sachkou}\ \emph {et~al.}(2019)\citenamefont
  {Sachkou}, \citenamefont {Baker}, \citenamefont {Harris}, \citenamefont
  {Stockdale}, \citenamefont {Forstner}, \citenamefont {Reeves}, \citenamefont
  {He}, \citenamefont {McAuslan}, \citenamefont {Bradley}, \citenamefont
  {Davis},\ and\ \citenamefont {Bowen}}]{sachkou2019coherent}%
  \BibitemOpen
  \bibfield  {author} {\bibinfo {author} {\bibfnamefont {Y.~P.}\ \bibnamefont
  {Sachkou}}, \bibinfo {author} {\bibfnamefont {C.~G.}\ \bibnamefont {Baker}},
  \bibinfo {author} {\bibfnamefont {G.~I.}\ \bibnamefont {Harris}}, \bibinfo
  {author} {\bibfnamefont {O.~R.}\ \bibnamefont {Stockdale}}, \bibinfo {author}
  {\bibfnamefont {S.}~\bibnamefont {Forstner}}, \bibinfo {author}
  {\bibfnamefont {M.~T.}\ \bibnamefont {Reeves}}, \bibinfo {author}
  {\bibfnamefont {X.}~\bibnamefont {He}}, \bibinfo {author} {\bibfnamefont
  {D.~L.}\ \bibnamefont {McAuslan}}, \bibinfo {author} {\bibfnamefont {A.~S.}\
  \bibnamefont {Bradley}}, \bibinfo {author} {\bibfnamefont {M.~J.}\
  \bibnamefont {Davis}},\ and\ \bibinfo {author} {\bibfnamefont {W.~P.}\
  \bibnamefont {Bowen}},\ }\bibfield  {title} {\bibinfo {title} {Coherent
  vortex dynamics in a strongly interacting superfluid on a silicon chip},\
  }\href@noop {} {\bibfield  {journal} {\bibinfo  {journal} {Science}\ }\textbf
  {\bibinfo {volume} {366}},\ \bibinfo {pages} {1480} (\bibinfo {year}
  {2019})}\BibitemShut {NoStop}%
\bibitem [{\citenamefont {Vagov}\ \emph {et~al.}(2016)\citenamefont {Vagov},
  \citenamefont {Shanenko}, \citenamefont {Milo{\v{s}}evi{\'c}}, \citenamefont
  {Axt}, \citenamefont {Vinokur}, \citenamefont {Aguiar},\ and\ \citenamefont
  {Peeters}}]{vagov2016superconductivity}%
  \BibitemOpen
  \bibfield  {author} {\bibinfo {author} {\bibfnamefont {A.}~\bibnamefont
  {Vagov}}, \bibinfo {author} {\bibfnamefont {A.~A.}\ \bibnamefont {Shanenko}},
  \bibinfo {author} {\bibfnamefont {M.~V.}\ \bibnamefont
  {Milo{\v{s}}evi{\'c}}}, \bibinfo {author} {\bibfnamefont {V.~M.}\
  \bibnamefont {Axt}}, \bibinfo {author} {\bibfnamefont {V.~M.}\ \bibnamefont
  {Vinokur}}, \bibinfo {author} {\bibfnamefont {J.~A.}\ \bibnamefont
  {Aguiar}},\ and\ \bibinfo {author} {\bibfnamefont {F.~M.}\ \bibnamefont
  {Peeters}},\ }\bibfield  {title} {\bibinfo {title} {Superconductivity between
  standard types: {Multiband} versus single-band materials},\ }\href@noop {}
  {\bibfield  {journal} {\bibinfo  {journal} {Phys. Rev. B}\ }\textbf {\bibinfo
  {volume} {93}},\ \bibinfo {pages} {174503} (\bibinfo {year}
  {2016})}\BibitemShut {NoStop}%
\bibitem [{\citenamefont {Aladyshkin}\ \emph {et~al.}(2009)\citenamefont
  {Aladyshkin}, \citenamefont {Silhanek}, \citenamefont {Gillijns},\ and\
  \citenamefont {Moshchalkov}}]{aladyshkin2009nucleation}%
  \BibitemOpen
  \bibfield  {author} {\bibinfo {author} {\bibfnamefont {A.~Y.}\ \bibnamefont
  {Aladyshkin}}, \bibinfo {author} {\bibfnamefont {A.~V.}\ \bibnamefont
  {Silhanek}}, \bibinfo {author} {\bibfnamefont {W.}~\bibnamefont {Gillijns}},\
  and\ \bibinfo {author} {\bibfnamefont {V.~V.}\ \bibnamefont {Moshchalkov}},\
  }\bibfield  {title} {\bibinfo {title} {Nucleation of superconductivity and
  vortex matter in superconductor--ferromagnethybrids},\ }\href@noop {}
  {\bibfield  {journal} {\bibinfo  {journal} {Supercond. Sci. Tech.}\ }\textbf
  {\bibinfo {volume} {22}},\ \bibinfo {pages} {053001} (\bibinfo {year}
  {2009})}\BibitemShut {NoStop}%
\bibitem [{\citenamefont {Stockdale}\ \emph {et~al.}(2020)\citenamefont
  {Stockdale}, \citenamefont {Reeves}, \citenamefont {Yu}, \citenamefont
  {Gauthier}, \citenamefont {Goddard-Lee}, \citenamefont {Bowen}, \citenamefont
  {Neely},\ and\ \citenamefont {Davis}}]{stockdale2020universal}%
  \BibitemOpen
  \bibfield  {author} {\bibinfo {author} {\bibfnamefont {O.~R.}\ \bibnamefont
  {Stockdale}}, \bibinfo {author} {\bibfnamefont {M.~T.}\ \bibnamefont
  {Reeves}}, \bibinfo {author} {\bibfnamefont {X.}~\bibnamefont {Yu}}, \bibinfo
  {author} {\bibfnamefont {G.}~\bibnamefont {Gauthier}}, \bibinfo {author}
  {\bibfnamefont {K.}~\bibnamefont {Goddard-Lee}}, \bibinfo {author}
  {\bibfnamefont {W.~P.}\ \bibnamefont {Bowen}}, \bibinfo {author}
  {\bibfnamefont {T.~W.}\ \bibnamefont {Neely}},\ and\ \bibinfo {author}
  {\bibfnamefont {M.~J.}\ \bibnamefont {Davis}},\ }\bibfield  {title} {\bibinfo
  {title} {Universal dynamics in the expansion of vortex clusters in a
  dissipative two-dimensional superfluid},\ }\href@noop {} {\bibfield
  {journal} {\bibinfo  {journal} {Phys. Rev. Research}\ }\textbf {\bibinfo
  {volume} {2}},\ \bibinfo {pages} {033138} (\bibinfo {year}
  {2020})}\BibitemShut {NoStop}%
\bibitem [{\citenamefont {Geelmuyden}\ \emph {et~al.}(2022)\citenamefont
  {Geelmuyden}, \citenamefont {Erne}, \citenamefont {Patrick}, \citenamefont
  {Barenghi},\ and\ \citenamefont {Weinfurtner}}]{geelmuyden2022sound}%
  \BibitemOpen
  \bibfield  {author} {\bibinfo {author} {\bibfnamefont {A.}~\bibnamefont
  {Geelmuyden}}, \bibinfo {author} {\bibfnamefont {S.}~\bibnamefont {Erne}},
  \bibinfo {author} {\bibfnamefont {S.}~\bibnamefont {Patrick}}, \bibinfo
  {author} {\bibfnamefont {C.~F.}\ \bibnamefont {Barenghi}},\ and\ \bibinfo
  {author} {\bibfnamefont {S.}~\bibnamefont {Weinfurtner}},\ }\bibfield
  {title} {\bibinfo {title} {Sound-ring radiation of expanding vortex
  clusters},\ }\href@noop {} {\bibfield  {journal} {\bibinfo  {journal} {Phys.
  Rev. Research}\ }\textbf {\bibinfo {volume} {4}},\ \bibinfo {pages} {023099}
  (\bibinfo {year} {2022})}\BibitemShut {NoStop}%
\bibitem [{\citenamefont {Cidrim}\ \emph {et~al.}(2017)\citenamefont {Cidrim},
  \citenamefont {White}, \citenamefont {Allen}, \citenamefont {Bagnato},\ and\
  \citenamefont {Barenghi}}]{cidrim2017vinen}%
  \BibitemOpen
  \bibfield  {author} {\bibinfo {author} {\bibfnamefont {A.}~\bibnamefont
  {Cidrim}}, \bibinfo {author} {\bibfnamefont {A.~C.}\ \bibnamefont {White}},
  \bibinfo {author} {\bibfnamefont {A.~J.}\ \bibnamefont {Allen}}, \bibinfo
  {author} {\bibfnamefont {V.~S.}\ \bibnamefont {Bagnato}},\ and\ \bibinfo
  {author} {\bibfnamefont {C.~F.}\ \bibnamefont {Barenghi}},\ }\bibfield
  {title} {\bibinfo {title} {Vinen turbulence via the decay of multicharged
  vortices in trapped atomic bose-einstein condensates},\ }\href@noop {}
  {\bibfield  {journal} {\bibinfo  {journal} {Phys. Rev. A}\ }\textbf {\bibinfo
  {volume} {96}},\ \bibinfo {pages} {023617} (\bibinfo {year}
  {2017})}\BibitemShut {NoStop}%
\bibitem [{\citenamefont {Chavanis}(2001)}]{chavanis2001kinetic}%
  \BibitemOpen
  \bibfield  {author} {\bibinfo {author} {\bibfnamefont {P.~H.}\ \bibnamefont
  {Chavanis}},\ }\bibfield  {title} {\bibinfo {title} {Kinetic theory of point
  vortices: diffusion coefficient and systematic drift},\ }\href@noop {}
  {\bibfield  {journal} {\bibinfo  {journal} {Phys. Rev. E}\ }\textbf {\bibinfo
  {volume} {64}},\ \bibinfo {pages} {026309} (\bibinfo {year}
  {2001})}\BibitemShut {NoStop}%
\bibitem [{\citenamefont {Chavanis}(2012)}]{chavanis2012kinetic2}%
  \BibitemOpen
  \bibfield  {author} {\bibinfo {author} {\bibfnamefont {P.~H.}\ \bibnamefont
  {Chavanis}},\ }\bibfield  {title} {\bibinfo {title} {Kinetic theory of
  onsager’s vortices in two-dimensional hydrodynamics},\ }\href@noop {}
  {\bibfield  {journal} {\bibinfo  {journal} {Physica A Stat. Mech. App.}\
  }\textbf {\bibinfo {volume} {391}},\ \bibinfo {pages} {3657} (\bibinfo {year}
  {2012})}\BibitemShut {NoStop}%
\bibitem [{\citenamefont {Sire}\ and\ \citenamefont
  {Chavanis}(2000)}]{sire2000numerical}%
  \BibitemOpen
  \bibfield  {author} {\bibinfo {author} {\bibfnamefont {C.}~\bibnamefont
  {Sire}}\ and\ \bibinfo {author} {\bibfnamefont {P.~H.}\ \bibnamefont
  {Chavanis}},\ }\bibfield  {title} {\bibinfo {title} {Numerical
  renormalization group of vortex aggregation in two-dimensional decaying
  turbulence: The role of three-body interactions},\ }\href@noop {} {\bibfield
  {journal} {\bibinfo  {journal} {Phys. Rev. E}\ }\textbf {\bibinfo {volume}
  {61}},\ \bibinfo {pages} {6644} (\bibinfo {year} {2000})}\BibitemShut
  {NoStop}%
\bibitem [{\citenamefont {Sire}\ \emph {et~al.}(2011)\citenamefont {Sire},
  \citenamefont {Chavanis},\ and\ \citenamefont {Sopik}}]{sire2011effective}%
  \BibitemOpen
  \bibfield  {author} {\bibinfo {author} {\bibfnamefont {C.}~\bibnamefont
  {Sire}}, \bibinfo {author} {\bibfnamefont {P.~H.}\ \bibnamefont {Chavanis}},\
  and\ \bibinfo {author} {\bibfnamefont {J.}~\bibnamefont {Sopik}},\ }\bibfield
   {title} {\bibinfo {title} {Effective merging dynamics of two and three fluid
  vortices: Application to two-dimensional decaying turbulence},\ }\href@noop
  {} {\bibfield  {journal} {\bibinfo  {journal} {Phys. Rev. E Stat. Nonlin.
  Soft Matter Phys.}\ }\textbf {\bibinfo {volume} {84}},\ \bibinfo {pages}
  {056317} (\bibinfo {year} {2011})}\BibitemShut {NoStop}%
\bibitem [{\citenamefont {Reeves}\ \emph {et~al.}(2013)\citenamefont {Reeves},
  \citenamefont {Billam}, \citenamefont {Anderson},\ and\ \citenamefont
  {Bradley}}]{reeves2013inverse}%
  \BibitemOpen
  \bibfield  {author} {\bibinfo {author} {\bibfnamefont {M.~T.}\ \bibnamefont
  {Reeves}}, \bibinfo {author} {\bibfnamefont {T.~P.}\ \bibnamefont {Billam}},
  \bibinfo {author} {\bibfnamefont {B.~P.}\ \bibnamefont {Anderson}},\ and\
  \bibinfo {author} {\bibfnamefont {A.~S.}\ \bibnamefont {Bradley}},\
  }\bibfield  {title} {\bibinfo {title} {Inverse energy cascade in forced
  two-dimensional quantum turbulence},\ }\href@noop {} {\bibfield  {journal}
  {\bibinfo  {journal} {Phys. Rev. Lett.}\ }\textbf {\bibinfo {volume} {110}},\
  \bibinfo {pages} {104501} (\bibinfo {year} {2013})}\BibitemShut {NoStop}%
\bibitem [{\citenamefont {Gauthier}\ \emph {et~al.}(2019)\citenamefont
  {Gauthier}, \citenamefont {Reeves}, \citenamefont {Yu}, \citenamefont
  {Bradley}, \citenamefont {Baker}, \citenamefont {Bell}, \citenamefont
  {Rubinsztein-Dunlop}, \citenamefont {Davis},\ and\ \citenamefont
  {Neely}}]{gauthier2019giant}%
  \BibitemOpen
  \bibfield  {author} {\bibinfo {author} {\bibfnamefont {G.}~\bibnamefont
  {Gauthier}}, \bibinfo {author} {\bibfnamefont {M.~T.}\ \bibnamefont
  {Reeves}}, \bibinfo {author} {\bibfnamefont {X.}~\bibnamefont {Yu}}, \bibinfo
  {author} {\bibfnamefont {A.~S.}\ \bibnamefont {Bradley}}, \bibinfo {author}
  {\bibfnamefont {M.~A.}\ \bibnamefont {Baker}}, \bibinfo {author}
  {\bibfnamefont {T.~A.}\ \bibnamefont {Bell}}, \bibinfo {author}
  {\bibfnamefont {H.}~\bibnamefont {Rubinsztein-Dunlop}}, \bibinfo {author}
  {\bibfnamefont {M.~J.}\ \bibnamefont {Davis}},\ and\ \bibinfo {author}
  {\bibfnamefont {T.~W.}\ \bibnamefont {Neely}},\ }\bibfield  {title} {\bibinfo
  {title} {Giant vortex clusters in a two-dimensional quantum fluid},\
  }\href@noop {} {\bibfield  {journal} {\bibinfo  {journal} {Science}\ }\textbf
  {\bibinfo {volume} {364}},\ \bibinfo {pages} {1264} (\bibinfo {year}
  {2019})}\BibitemShut {NoStop}%
\bibitem [{\citenamefont {Bradley}\ and\ \citenamefont
  {Anderson}(2012)}]{bradley2012energy}%
  \BibitemOpen
  \bibfield  {author} {\bibinfo {author} {\bibfnamefont {A.~S.}\ \bibnamefont
  {Bradley}}\ and\ \bibinfo {author} {\bibfnamefont {B.~P.}\ \bibnamefont
  {Anderson}},\ }\bibfield  {title} {\bibinfo {title} {Energy spectra of vortex
  distributions in two-dimensional quantum turbulence},\ }\href@noop {}
  {\bibfield  {journal} {\bibinfo  {journal} {Physical Review X}\ }\textbf
  {\bibinfo {volume} {2}},\ \bibinfo {pages} {041001} (\bibinfo {year}
  {2012})}\BibitemShut {NoStop}%
\bibitem [{\citenamefont {Yu}\ \emph {et~al.}(2016)\citenamefont {Yu},
  \citenamefont {Billam}, \citenamefont {Nian}, \citenamefont {Reeves},\ and\
  \citenamefont {Bradley}}]{yu2016theory}%
  \BibitemOpen
  \bibfield  {author} {\bibinfo {author} {\bibfnamefont {X.}~\bibnamefont
  {Yu}}, \bibinfo {author} {\bibfnamefont {T.~P.}\ \bibnamefont {Billam}},
  \bibinfo {author} {\bibfnamefont {J.}~\bibnamefont {Nian}}, \bibinfo {author}
  {\bibfnamefont {M.~T.}\ \bibnamefont {Reeves}},\ and\ \bibinfo {author}
  {\bibfnamefont {A.~S.}\ \bibnamefont {Bradley}},\ }\bibfield  {title}
  {\bibinfo {title} {Theory of the vortex-clustering transition in a confined
  two-dimensional quantum fluid},\ }\href@noop {} {\bibfield  {journal}
  {\bibinfo  {journal} {Phys. Rev. A}\ }\textbf {\bibinfo {volume} {94}},\
  \bibinfo {pages} {023602} (\bibinfo {year} {2016})}\BibitemShut {NoStop}%
\end{thebibliography}%
\end{document}